\documentclass{article}
\usepackage[english]{babel} 
\usepackage{mathdots}
\usepackage{amsmath}
\usepackage{amssymb}
\usepackage{mathtools}
\usepackage{multirow}
\usepackage{graphicx}
\usepackage{epstopdf}
\usepackage{inputenc}
\usepackage{geometry}
\usepackage{authblk}
\usepackage{hyperref}
\providecommand{\keywords}[1]{\textbf{\textit{Key Words and Phrases:}} #1}
\numberwithin{equation}{section}
\DeclareUnicodeCharacter{2212}{\textendash}

\begin{document}

\title{\textbf{Thermal oscillations and resonance in electron-phonon interaction process } }

\author[*,1,4]{\textbf{Emad Awad}}
\author[$\dagger$,2]{\textbf{Weizhong Dai}}
\author[$\ddagger$,3]{\textbf{Sergey Sobolev }}
\affil[1]{Department of Mathematics, Faculty of Education, Alexandria University, Souter St. El-Shatby, Alexandria P.O. box 21526, Egypt.}
\affil[2]{Mathematics \& Statistics, College of Engineering \& Science, Louisiana Tech University, Ruston, LA 71272, USA.}
\affil[3]{Institute of Problems of Chemical Physics, Academy of Sciences of Russia, Chernogolovka, Moscow Region 142432, Russia.}
\affil[*]{E-mail Address: emadawad78@alexu.edu.eg}
\affil[$\dagger$]{E-mail Address: dai@coes.latech.edu}
\affil[$\ddagger$]{E-mail Address: sobolev@icp.ac.ru}
\affil[4]{Author to whom any correspondence should be addressed}
\renewcommand\Authands{ and }

\maketitle

\begin{abstract}
	A recent theoretical study (Xu, Proc. R. Soc. A: Math. Phys. Eng. Sci., 2021) has derived conditions on the coefficients of Jeffreys-type equation to predict thermal oscillations and resonance during phonon hydrodynamics in non-metallic solids. Thermal resonance, in which the temperature amplitude attains a maximum value (peak) in response to an external exciting frequency source, is a phenomenon pertinent to the presence of underdamped thermal oscillations and explicit finite-speed for the thermal wave propagation. The present work investigates the occurrence condition for thermal resonance phenomenon during the electron-phonon interaction process in metals based on the hyperbolic two-temperature model. First, a sufficient condition for underdamped electron and lattice temperature oscillations is discussed by deriving a critical frequency (a material characteristic). It is shown that the critical frequency of thermal waves near room temperature, during electron-phonon interactions, may be on the order of terahertz ($10-20$ THz for Cu and Au, i.e., lying within the terahertz gap). It is found that whenever the natural frequency of metal temperature exceeds this frequency threshold, the temperature oscillations are of underdamped type. However, this condition is not necessary, since there is a small frequency domain, below this threshold, in which the underdamped thermal wave solution is available but not effective. Otherwise, the critical damping and the overdamping conditions of the temperature waves are determined numerically for a sample of pure metals. The thermal resonance conditions in both electron and lattice temperatures are investigated. The occurrence of resonance in both electron and lattice temperature is conditional on violating two distinct critical values of frequencies. When the natural frequency of the system becomes larger than these two critical values, an applied frequency equal to such a natural frequency can drive both electron and lattice temperatures to resonate together with different amplitudes and behaviors. However, the electron temperature resonates earlier than the lattice temperature.
\end{abstract}

\tableofcontents

\section{ Introduction}

The wavy nature of various dynamical subsystems, e.g., photons, phonons, and electrons, is one of the well-known scientific universals \cite{RN1,RN2}. It is unreasonable to reject the idea that thermal waves propagate at a finite speed, even in cases when it is not technically feasible to quantify the speed of thermal waves in a solid metallic medium. During heating of metal films by a high exciting frequency heat source, it is presumed that the electron subsystem is responsible to transfer energy from photons to the metal lattice \cite{RN3,RN4,RN5}. This process was mathematically described through the two-step (or two-temperature) model, which assigns two distinct temperatures to the electrons and the lattice during the transient response of metals to ultrafast heat source \cite{RN6,RN7,RN8,RN9,RN10,RN11,RN12,RN13}. The presence of multi-energy-carriers of the thermal system leads to the multi-temperature postulate not only two temperatures \cite{RN14,RN15,RN16}. In the absence of free electrons, as the case of non-metallic (insulators) solids, the lattice vibration (phonon) is the solely responsible for heat transfer \cite{RN17}. The classical thermal wave model allows a single step (one-step) of interactions, precisely, absorption of thermal energy by electrons or by lattice in non-metallic solids \cite{RN8,RN17}. 

The one-step model of heat conduction, known in literature as the classical thermal damped wave model or the Maxwell-Cattaneo-Vernotte (MCV) equation, was theoretically founded and examined in many past works, see e.g., extended irreversible thermodynamics \cite{RN18}. It is characterized by the presence of a relaxation time that results in a finite speed for thermal wave propagation, where the absence of such a time constant leads to the infinite speed characterizing thermal diffusion. From an experimental point of view, the finite speed of thermal wave (measurable value) was observed in many materials, e.g., in liquid helium \cite{RN19,RN20}, solid helium \cite{RN21,RN22}, in sodium fluoride \cite{RN23,RN24}, in bismuth \cite{RN25}, and most recently, it is observed in graphite at low temperature (above $100\ \mathrm{K}$) \cite{RN26}. It is worthy referring  to the possibility of activating the electron-phonon interaction and thermal relaxation in graphite ultrathin films \cite{RN27,RN28}. On the other hand, because of the high value of thermal conductivity in pure metals, the thermal waves propagate very fast within the conductor (i.e., $v\to \infty $, or ${1}/{v}\to 0$). Although the wide applicability of the one-step model of heat conduction, it could not simulate Brorson et al experiment of ultrashort laser heating of gold film \cite{RN8,RN29}. In contrast to the one-step heat conduction model, the phenomenological parabolic two-step model can reproduce numerical data agree well with the experimental results of ultrafast heating of metals \cite{RN8}. Being owing a parabolic partial differential equation governing both electron and lattice temperatures, the phenomenological two-step model results in a diffusive nature for the thermal waves resulting from the absence of a definite value for the thermal waves speed. An attempt suggesting a hyperbolic two temperatures model, with two relaxation times and thus with two distinct speeds (the Fermi velocity $v_F$ and the sound velocity in metals $v_c$), was due to Sobolev \cite{RN6}. A similar model containing two relaxation times for the electron and the lattice heat fluxes was numerically discussed thereafter by Chen and Beraun \cite{RN10}. The two-temperature model shows good agreements with experimental results in the multilayers metals, compared with the one-step (damped wave) model \cite{RN30,RN31}, see also recent numerical experiments on the two-temperature models \cite{RN32,RN33} and a possible connection to the macroscopic law in \cite{RN34}.

The thermal resonance phenomenon, in which the temperature amplitude can be amplified at a certain frequency of the external applied heat source, is a conditional phenomenon pertinent not only to the oscillations of temperature and the finite thermal wave speed, but also the verification of underdamping for temperature oscillations and a relationship between the external applied frequency and the natural frequency of thermal wave \cite{RN35}. Such an amplification in the temperature, or specifically the aggregation in the time-dependent amplitude of temperature may be a compelling reason for the thermal damage. Over than two decades ago, a contribution by Xu and Wang proved that the parabolic dual-phase-lag heat conduction law may exhibit underdamped temperature oscillations and thermal resonance in the case of temperature-gradient precedence, namely, the telegrapher-like case \cite{RN36}. Most recently, a Jeffreys-type heat conduction equation, similar to the dual-phase-lag model with different phase lag constants related to the Guyer-Krumhansl model, with different heat source structure has been investigated for underdamped oscillations and thermal resonance in the higher dimensions \cite{RN37}. 

The goal of the present work is to investigate the sufficient conditions for underdamped temperature oscillations and the possibility of thermal resonance occurrence in both the electron and the lattice temperatures. Because of the mathematical complexity of the problem, such as the presence of biquartic and bicubic equations, some conditions cannot be extracted analytically, rather we adopt an inductive numerical-based method depending on the symbolic calculations of the roots and deriving a general condition. The work is organized as follows: In the following section, \S\ \ref{sec2}, we outline the mathematical description of the parabolic and hyperbolic two-step models of heat transfer in metals. In \S\ \ref{sec3}, we derive the sufficient condition for the underdamped temperature oscillations and define the critical frequency at which the temperature oscillations are critically damped. The thermal resonance phenomenon of electron and lattice temperature is explored throughout section \ref{sec4}. Lastly, we summarize the results of the work and draw our conclusions in \S\ \ref{sec5}.

\section{ Hyperbolic two-step model}
\label{sec2}

The phenomenological parabolic two-step model describing the electron and the lattice temperatures during electron-phonon interactions in metals is defined through the system \cite{RN5,RN6,RN8,RN11}:

\begin{equation}\label{eq1}
C_e\frac{\partial T_e\left({\mathbf{r}},t\right)}{\partial t}=-\mathrm{\nabla }\cdot {\mathbf{q}}\left({\mathbf{r}},t\right)-G\left[T_e\left({\mathbf{r}},t\right)-T_l\left({\mathbf{r}},t\right)\right]+Q\left({\mathbf{r}},t\right),
\end{equation}
\begin{equation}\label{eq2}
C_l\frac{\partial T_l\left({\mathbf{r}},t\right)}{\partial t}=G\left[T_e\left({\mathbf{r}},t\right)-T_l\left({\mathbf{r}},t\right)\right],
\end{equation}
\begin{equation}\label{eq3}
{\mathbf{q}}\left({\mathbf{r}},t\right)=-k_e\mathrm{\nabla }T_e\left({\mathbf{r}},t\right),
\end{equation}
where $C_e$ and $C_l$ are respectively the heat capacities of the electron gas and the metal lattice, $G$ is the electron-phonon coupling factor, and $k_e$ is the thermal conductivity of electron subsystem. In addition, $T_e\left({\mathbf{r}},t\right)$ and $T_l\left({\mathbf{r}},t\right)$ denote respectively to the electron and lattice temperature variations measured from the room temperature $T_0$, namely, $T_e\left({\mathbf{r}},t\right)={\check{T}}_e\left({\mathbf{r}},t\right)-T_0$ and $T_l\left({\mathbf{r}},t\right)={\check{T}}_l\left({\mathbf{r}},t\right)-T_0$, where ${\check{T}}_e\left({\mathbf{r}},t\right)$ and ${\check{T}}_l\left({\mathbf{r}},t\right)$ are the absolute electron and lattice temperatures. The heat flux vector is denoted by ${\mathbf{q}}\left({\mathbf{r}},t\right)$, where ${\mathbf{r}}$ and $t$ are the position vector ${\mathbf{r}}{\mathbf{=}}\left\langle x,y,z\right\rangle $ and the time, and $\mathrm{\nabla }$ is the gradient operator. Lastly, the term $Q\left({\mathbf{r}},t\right)$ refers to the external heat source term. The model coefficients $C_e$, $C_l$, $G$ and $k_e$ are in general functions of space, however, we consider in our analysis a homogeneous heat conductor with temperature changes near the room temperature so that the material properties can be considered as constants measured near the room temperature. Upon eliminating the electron heat flux ${\mathbf{q}}\left({\mathbf{r}},t\right)$ among \eqref{eq1}-\eqref{eq3}, the energy equations governing the electron and lattice temperatures are given by \cite{RN6,RN38}

\begin{equation}\label{eq4}
\left(\frac{C_e+C_l}{k_e}\frac{\partial }{\partial t}+\frac{C_eC_l}{Gk_e}\frac{{\partial }^2}{\partial t^2}\right)\left[ \begin{array}{c}
T_e \\ 
T_l \end{array}
\right]=\left(1+\frac{C_l}{G}\frac{\partial }{\partial t}\right)\mathrm{\Delta }\left[ \begin{array}{c}
T_e \\ 
T_l \end{array}
\right]+\frac{1}{k_e}\left[ \begin{array}{c}
\left(1+\frac{C_l}{G}\frac{\partial }{\partial t}\right)Q \\ 
Q \end{array}
\right],
\end{equation}
where $\mathrm{\Delta }\mathrm{=}\mathrm{\nabla }\mathrm{\cdot }\mathrm{\nabla }$ is the Laplace operator. Equation \eqref{eq4} reduces to the classical Fourier conduction law when the lattice heat capacity is neglected, i.e., $C_l\ll G$, or ${C_l}/{G}\to 0$. Then, both lattice temperature and electron temperature are the same. Otherwise, equation \eqref{eq4} is a parabolic-type partial differential equation, representing a thermal wave propagating with infinite speed. When $Q=0$, the non-negativity of the Green functions of the resulting partial differential equation has been studied in \cite{RN39}, and it has been shown to be the space-time limit of a decoupled continuous-time random walk scheme \cite{RN40}. The underdamping characteristic is absent from equation \eqref{eq4}, with the virtue that the lattice heat capacity is larger than the electron heat capacity, $C_l\gg C_e$. Indeed, rearranging equation \eqref{eq4} to be written as \cite{RN41}

\begin{equation}\label{eq5}
\frac{C_e+C_l}{k_e}\left\{1+\frac{C_l}{G}{\left(1+\frac{C_l}{C_e}\right)}^{-1}\frac{\partial }{\partial t}\right\}\frac{\partial }{\partial t}\left[ \begin{array}{c}
T_e \\ 
T_l \end{array}
\right]=\left(1+\frac{C_l}{G}\frac{\partial }{\partial t}\right)\mathrm{\Delta }\left[ \begin{array}{c}
T_e \\ 
T_l \end{array}
\right]+\frac{1}{k_e}\left[ \begin{array}{c}
\left(1+\frac{C_l}{G}\frac{\partial }{\partial t}\right)Q \\ 
Q \end{array}
\right],
\end{equation}
and comparing between the coefficients of first-order time derivative in both sides, we have that $\frac{C_l}{G}{\left(1+\frac{C_l}{C_e}\right)}^{-1}\ll \frac{C_l}{G}$, according to the fact $C_l\gg C_e$, hence the absence of underdamping oscillations and thermal resonance \cite{RN36}. In \cite{RN6,RN9}, the authors considered modifying the constitutive law \eqref{eq3} to the form

\begin{equation} \label{eq6}
{\mathbf{q}}\left({\mathbf{r}},t\right)+{\tau }_F\frac{\partial {\mathbf{q}}\left({\mathbf{r}},t\right)}{\partial t}=-k_e\mathrm{\nabla }T_e\left({\mathbf{r}},t\right),
\end{equation}
where ${\tau }_F$ is the relaxation time evaluated at the Fermi surface. When the heat flux is eliminated among equations \eqref{eq1}, \eqref{eq2} and \eqref{eq6}, we obtain energy equations governing the electron and lattice temperature on the form \cite{RN13,RN16}:

\begin{multline} \label{eq7}
\frac{C_e+C_l}{k_e}\left\{\frac{\partial }{\partial t}+\left({\tau }_F+\frac{C_eC_l}{G\left(C_e+C_l\right)}\right)\frac{{\partial }^2}{\partial t^2}+\frac{C_eC_l{\tau }_F}{G\left(C_e+C_l\right)}\frac{{\partial }^3}{\partial t^3}\right\}\left[ \begin{array}{c}
T_e \\ 
T_l \end{array}
\right]=\\
\left(1+\frac{C_l}{G}\frac{\partial }{\partial t}\right)\mathrm{\Delta }\left[ \begin{array}{c}
T_e \\ 
T_l \end{array}
\right]+\frac{1}{k_e}\left[ \begin{array}{c}
\left(1+{\tau }_F\frac{\partial }{\partial t}\right)\left(1+\frac{C_l}{G}\frac{\partial }{\partial t}\right)Q \\ 
\left(1+{\tau }_F\frac{\partial }{\partial t}\right)Q \end{array}
\right].
\end{multline}

Therefore, the insertion of the thermal relaxation ${\tau }_F$ causes the presence of a third-order time derivative in the energy equation \eqref{eq7}, and thereby a thermal wave with definite speed of propagation $v_{HTS}$ given by \cite{RN34}:

\begin{equation}\label{eq8}
v_{HTS}=\sqrt{\frac{k_e}{{\tau }_FC_e}}.
\end{equation}
It is salient that when ${\tau }_F\to 0$, the hyperbolic-type energy equation \eqref{eq7} reduces to the parabolic-type energy equation \eqref{eq4} and the speed of thermal waves \eqref{eq8} becomes infinite, $v\to \infty $. 

For the sake of simplifying the mathematical analysis in the rest of this work, we introduce the following quantities \cite{RN34,RN41}:

\begin{equation}\label{eq9}
{\tau }_{q1}={\tau }_F,\ \ \ {\tau }_{q2}=\frac{C_eC_l}{G\left(C_e+C_l\right)},\ \ \ {\tau }_T=\frac{C_l}{G},\ \ \ \alpha =\frac{k_e}{C_e+C_l}.
\end{equation}

Utilizing the quantities \eqref{eq9}, the hyperbolic-type energy equations governing the electron and the lattice temperatures \eqref{eq7} can be written as

\begin{equation}\label{eq10}
\left\{\frac{1}{\alpha }\frac{\partial }{\partial t}+\left(\frac{1}{v^2_1}+\frac{1}{v^2_2}\right)\frac{{\partial }^2}{\partial t^2}+\frac{{\tau }_T}{v^2_3}\frac{{\partial }^3}{\partial t^3}\right\}\left[ \begin{array}{c}
T_e \\ 
T_l \end{array}
\right]=\left(1+{\tau }_T\frac{\partial }{\partial t}\right)\mathrm{\Delta }\left[ \begin{array}{c}
T_e \\ 
T_l \end{array}
\right]+\frac{1}{k_e}\left[ \begin{array}{c}
\left(1+{\tau }_{q1}\frac{\partial }{\partial t}\right)\left(1+{\tau }_T\frac{\partial }{\partial t}\right)Q \\ 
\left(1+{\tau }_{q1}\frac{\partial }{\partial t}\right)Q \end{array}
\right],
\end{equation}
where $v_1$, $v_2$ and $v_3$ are three material constants of the velocity dimension $\left[\mathrm{L}{\mathrm{T}}^{\mathrm{-}\mathrm{1}}\right]$, determined through

\begin{equation}\label{eq11}
v_1=\sqrt{\frac{\alpha }{{\tau }_{q1}}}=\sqrt{\frac{k_e}{\left(C_e+C_l\right){\tau }_F}},\ \ \ v_2=\sqrt{\frac{\alpha }{{\tau }_{q2}}}=\sqrt{\frac{Gk_e}{C_eC_l}},\ \ \ v_3=\sqrt{\frac{\alpha {\tau }_T}{{\tau }_{q1}{\tau }_{q2}}}=\sqrt{\frac{k_e}{{\tau }_FC_e}}=v_{HTS}.
\end{equation}

Furthermore, when the lattice heat capacity is negligible compared to the coupling factor, i.e., $C_l\ll G$, then ${C_l}/{G}\to 0$ or ${\tau }_T\to 0$ and ${\tau }_{q2}\to 0$, the energy equation \eqref{eq10} reduces to 

\begin{equation}\label{eq12}
\left\{\frac{1}{\alpha }\frac{\partial }{\partial t}+\frac{1}{v^2_1}\frac{{\partial }^2}{\partial t^2}\right\}\left[ \begin{array}{c}
T_e \\ 
T_l \end{array}
\right]=\mathrm{\Delta }\left[ \begin{array}{c}
T_e \\ 
T_l \end{array}
\right]+\frac{1}{k_e}\left[ \begin{array}{c}
\left(1+{\tau }_{q1}\frac{\partial }{\partial t}\right)Q \\ 
\left(1+{\tau }_{q1}\frac{\partial }{\partial t}\right)Q \end{array}
\right],
\end{equation}
namely, the coincidence of both electron and lattice temperatures. Equation \eqref{eq12} is the well-known classical damped wave equation.

\section{ Underdamped electron and lattice temperature oscillations  }
\label{sec3}
In this section, we find a sufficient condition for the thermal wave governed by the third-order partial differential equation \eqref{eq10} to produce temperature oscillations of the underdamped type and define for specific sample of metals the frequency domain in which the temperature oscillations are overdamped. Because the heat source term does not affect the nature of thermal waves, we focus our attention on the following form of partial differential equations:

\begin{equation}\label{eq13}
\frac{1}{\alpha }\frac{\partial T}{\partial t}+\left(\frac{1}{v^2_1}+\frac{1}{v^2_2}\right)\frac{{\partial }^2T}{\partial t^2}+\frac{{\tau }_T}{v^2_3}\frac{{\partial }^3T}{\partial t^3}=\left(1+{\tau }_T\frac{\partial }{\partial t}\right)\frac{{\partial }^2T}{\partial x^2},
\end{equation}
where the one-dimensional setting is considered, and $T$ is the temperature variation measured from the room temperature $T_0$, i.e., $T=\check{T}-T_0$ and $\check{T}$ is the absolute temperature. There are three characteristic speeds controlling equation \eqref{eq13}; $v_1$, $v_2$ and $v_3$ which satisfy the following inequalities: 

\begin{equation}\label{eq14}
v_3>v_1,\ \ \frac{1}{v^2_3}<\frac{1}{v^2_1}+\frac{1}{v^2_2}.
\end{equation}
In fact, the first inequality can be validated by the virtue of $C_l\gg C_e$. Indeed, we have that $v^2_3=\left(\frac{\alpha }{{\tau }_{q1}}\right)\left(\frac{{\tau }_T}{{\tau }_{q2}}\right)=v^2_1\left(1+\frac{C_l}{C_e}\right)$, which leads to $v^2_3>v^2_1$ or $v_3>v_1$. Lastly, the inequality $v^2_3>v^2_1$ yields $\frac{1}{v^2_3}<\frac{1}{v^2_1}<\frac{1}{v^2_1}+\frac{1}{v^2_2}$, hence the proof of inequalities \eqref{eq14}. Moreover, the numerical values for Gold (Au), Copper (Cu), and Lead (Pb), suggest generalizing \eqref{eq14} to the inequality (see Table \ref{tabel1}):
\begin{equation} \label{eq15}
	v_3>v_1>v_2.
\end{equation}

	\begin{table}
		\centering
		\caption{Material properties for Copper (Cu), Gold (Au), and Lead (Pb) at \textit{$T_0=300\ \mathrm{K}$}, refer to\cite{RN9,RN41}.}
		\label{tabel1}
\scalebox{0.83}{
	\begin{tabular}{||cccccccccccc||} 
		\hline\hline
		\text{}& $C_e $& $C_l $ & $G $ & $k_e $ &$\alpha$  & $\tau_F=\tau_{q1} $ & $\tau_{q2} $ & $\tau_T $ & $v_1 $ &$v_2 $  & $v_3 $  \\ 
		{\textbf{Metal}} & $\mathrm{Jm^{-3}K^{-1}}$ & $\mathrm{Jm^{-3}K^{-1}}$ & $\mathrm{Wm^{-3}K^{-1}}$ & $\mathrm{Wm^{-1}K^{-1}}$ & $\mathrm{m^{2}s^{-1}}$ & Picosecond & Picosecond & Picosecond &  $\mathrm{ms^{-1}}$&$\mathrm{ms^{-1}}$  &$\mathrm{ms^{-1}}$  \\ 
		\text{}& $\times 10^{4}$ & $\times 10^{6}$  & $\times 10^{16}$ &  &$\times 10^{-4}$  & $\mathrm{(ps)}$  & $\mathrm{(ps)}$  & $\mathrm{(ps)}$ & $\times 10^{4}$ & $\times 10^{4}$ &$\times 10^{4}$  \\ 
		\hline\hline
		{\textbf{Cu}}&2.1  & 3.4 & 4.8 & 368 & 1.076 & 0.03 &0.4348  & 70.33 & 5.988 &1.573  & 76.16 \\ \hline 
		{\textbf{Au}}&2.1 &2.5  & 2.8 & 315 & 1.25 & 0.04 &0.7438  &89.28  &  5.589&1.296  &61.23  \\ \hline 
		{\textbf{Pb}}&2.1  &1.5  & 12.4 & 35 & 0.2301 &0.005  &0.167  &12.097  &6.784  &1.174  &57.74  \\ 
		\hline\hline
	\end{tabular}
}
	\end{table}
In \eqref{eq13}, the term ${{\tau }_T}/{v^2_3}$ is a very small number, about $23.8\times {10}^{-23}$ for Gold near room temperature $T_0=300\ \mathrm{K}$. To activate this velocity $v_3$, the response time of metal should be on an order of magnitude sufficiently smaller than ${\tau }_T$ such that the term $\frac{{\tau }_T}{v^2_3t^3}$ becomes sufficiently larger than other terms of the left-hand-side, $\frac{1}{v^2_1t^2}$, $\frac{1}{v^2_2t^2}$, and $\frac{1}{\alpha t}$, hence the second inequality of \eqref{eq14} is reversed. For example, at a response time $400$ attoseconds ($=400\times {10}^{-18}\ \mathrm{s}$), namely, it is larger than the time needed to electrons for hopping among atoms (320 attoseconds \cite{RN42}) and it is of order smaller than the order of ${\tau }_T$ (${10}^{-14}$ ), we can confidently write

\begin{equation} \label{eq16}
\frac{{\tau }_T}{v^2_3t^3}\gg \frac{1}{v^2_1t^2}+\frac{1}{v^2_2t^2}\gg \frac{1}{\alpha t},\ \ \ \frac{{\tau }_T}{t}\gg 1,
\end{equation}
which indicates that the term containing the third-order time derivative on the left-hand-side of \eqref{eq13} and the term containing the first order time derivative on the right-hand-side of \eqref{eq13} dominate this short-time (attosecond) domain. As the response time prolongs to 100 picoseconds ($=1\times {10}^{-10}\ \mathrm{s}$), the third term on the left-hand-side of \eqref{eq13} becomes negligible compared to the other terms, i.e.,

\begin{equation} \label{eq17}
\frac{{\tau }_T}{v^2_3t^3}\ll \frac{1}{v^2_1t^2}+\frac{1}{v^2_2t^2}+\frac{1}{\alpha t},\qquad\frac{{\tau }_T}{t}\ll 1,
\end{equation}
namely, equation \eqref{eq13} behaves then like the classical damped wave equation \eqref{eq12}. Therefore, we can state that in the attosecond domain equation \eqref{eq13} behaves like a thermal undamped wave propagating with velocity $v_3$:

\begin{equation} \label{eq18}
\frac{1}{v^2_3}\frac{{\partial }^2T}{\partial t^2}=\frac{{\partial }^2T}{\partial x^2}.
\end{equation}

Let ${\omega }_n$ be the natural frequency of the thermal wave in the short time domain, then the solution of \eqref{eq18} can be written on the form:

\begin{equation} \label{eq19}
 T\left(x,t\right)\coloneqq \sum^{\infty }_{n=1}{{\phi }_n\left(x\right){\exp \left(\imath {\omega }_nt\right)\ }},
\end{equation}
where ${\phi }_n\left(x\right)$ are eigenfunctions to be determined. Furthermore, we consider the boundary value problem of a thermal wave propagating in a one-dimensional thick plate of thickness $L$, governed by the wave equation \eqref{eq18} and subject to the boundary conditions:

\begin{equation} \label{eq20}
 T\left(0,t\right)=T\left(L,t\right)=0,
\end{equation}
i.e., the plate surfaces are kept at room temperature, ($\check{T}=T_0$). Upon using the proposed modal representation \eqref{eq19} into \eqref{eq18} we obtain

\begin{equation} \label{eq21}
\frac{{\mathrm{d}}^{\mathrm{2}}{\phi }_n\left(x\right)}{\mathrm{d}x^2}=-{\left(\frac{{\omega }_n}{v_3}\right)}^2{\phi }_n\left(x\right).
\end{equation}

The solution of \eqref{eq21} can be written on the form ${\phi }_n\left(x\right)=a_n{\cos \left(\frac{{\omega }_nx}{v_3}\right)\ }+b_n{\sin \left(\frac{{\omega }_nx}{v_3}\right)\ }$. In order to determine exactly the set of eigenfunctions  $\{{\phi }_n\left(x\right)\}$, we utilize the boundary conditions \eqref{eq20} which suggest that $a_n=0$ and $b_n{\sin \left(\frac{{\omega }_nL}{v_3}\right)\ }=0$. Since $b_n\neq 0$, then the only possibility for the second condition of \eqref{eq20} is that 

\begin{equation} \label{eq22}
{\omega }_n=\frac{n\pi v_3}{L},
\end{equation}
which provide the set of orthogonal eigenfunctions on the form

\begin{equation} \label{eq23}
{\phi }_n\left(x\right)\in \left\{{\sin \left(\frac{n\pi x}{L}\right)\ },\ \ \ n=1,2,\dots \right\}=\left\{{\sin \left(\frac{\pi x}{L}\right)\ },\ {\sin \left(\frac{2\pi x}{L}\right)\ },\dots \right\}.
\end{equation}

Returning to the general equation governing the thermal waves during electron-phonon interaction \eqref{eq13}, instead of the special wave solution \eqref{eq19} valid for specific time domain in which the temperature oscillates as a wave with natural frequency ${\omega }_n$, we now suggest the general representation valid along the whole timeline:

\begin{equation} \label{eq24}
T\left(x,t\right)\coloneqq \sum^{\infty }_{n=1}{{\mathrm{\Gamma }}_n\left(t\right){\phi }_n\left(x\right)},
\end{equation}
where ${\mathrm{\Gamma }}_n\left(t\right)$ is the time-dependent amplitude of temperature and ${\phi }_n\left(x\right)$ is on of the orthogonal eigenfunction set \eqref{eq23}. Upon inserting the general representation \eqref{eq24} into \eqref{eq13}, we obtain 

\begin{equation} \label{eq25}
\frac{{\mathrm{d}}^3{\mathrm{\Gamma }}_n\left(t\right)}{\mathrm{d}t^3}+\frac{1}{{\tau }_T}\left(\frac{v^2_3}{v^2_1}+\frac{v^2_3}{v^2_2}\right)\frac{{\mathrm{d}}^2{\mathrm{\Gamma }}_n\left(t\right)}{\mathrm{d}t^2}+\left(\frac{v^2_3}{\alpha {\tau }_T}+{\omega }^2_n\right)\frac{\mathrm{d}{\mathrm{\Gamma }}_n\left(t\right)}{\mathrm{d}t}+\frac{{\omega }^2_n}{{\tau }_T}{\mathrm{\Gamma }}_n\left(t\right)=0.
\end{equation}

To explore the features of thermal waves governed by the energy equation \eqref{eq13}, it is reasoning to examine the properties of the time-dependent amplitude ${\mathrm{\Gamma }}_n\left(t\right)$ governed by \eqref{eq25}. The solution of \eqref{eq25} is ${\mathrm{\Gamma }}_n\left(t\right)={\exp \left(\lambda t\right)\ }$, where $\lambda $ is determined through

\begin{equation} \label{eq26}
{\lambda }^3+\frac{1}{{\tau }_T}\left(\frac{v^2_3}{v^2_1}+\frac{v^2_3}{v^2_2}\right){\lambda }^2+\left(\frac{v^2_3}{\alpha {\tau }_T}+{\omega }^2_n\right)\lambda +\frac{{\omega }^2_n}{{\tau }_T}=0.
\end{equation}

The cubic characteristic equation \eqref{eq26} includes two cases: (i) The classical damped wave; and (ii) The Two-Step damped wave model.

\subsection{ Classical damped wave solution}

\noindent In the case of disregarding the heat capacity of the metal lattice, i.e., ${C_l}/{G}\to 0$ or ${\tau }_T\to 0$, ${\tau }_{q2}\to 0$, ${{\tau }_T}/{{\tau }_{q2}}\to $1, and $v_3\to v_1$, equation \eqref{eq13} reduces to the classical damped wave equation:

\begin{equation} \label{eq27}
\frac{1}{\alpha }\frac{\partial T}{\partial t}+\frac{1}{v^2_1}\frac{{\partial }^2T}{\partial t^2}=\frac{{\partial }^2T}{\partial x^2},
\end{equation}
and the characteristic equation \eqref{eq26} reduces to \cite{RN35}

\begin{equation} \label{eq28}
{\lambda }^2+\frac{v^2_1}{\alpha }\lambda +{\omega }^2_n=0,
\end{equation}
which has the following roots

\begin{equation} \label{eq29}
{\lambda }_{1,2}=-\frac{1}{2{\tau }_{q1}}\pm \sqrt{\frac{1}{4{\tau }^2_{q1}}-{\omega }^2_n}.
\end{equation}

In view of equation \eqref{eq29}, we have three different solutions depending on the value of natural frequency.:

\begin{enumerate}
\item[(i)] When  ${\omega }_n<{1}/{(2{\tau }_{q1})}$, we have two distinct real roots, and the solution is given as

\begin{equation} \label{eq30}
{\mathrm{\Gamma }}_n\left(t\right)={\exp \left(-\frac{t}{2{\tau }_{q1}}\right)\ }\left[A{\exp \left(\frac{t}{2{\tau }_{q1}}\sqrt{1-4{\tau }^2_{q1}{\omega }^2_n}\right)\ }+B{\exp \left(-\frac{t}{2{\tau }_{q1}}\sqrt{1-4{\tau }^2_{q1}{\omega }^2_n}\right)\ }\right].
\end{equation}
This case is called the overdamped temperature oscillations.

\item[(ii)]  When ${\omega }_n={1}/{(2{\tau }_{q1})}$, we have two equal real roots ${\lambda }_1={\lambda }_1=-{1}/{\left(2{\tau }_{q1}\right)}$, and the solution is given as

\begin{equation} \label{eq31}
{\mathrm{\Gamma }}_n\left(t\right)={\exp \left(-\frac{t}{2{\tau }_{q1}}\right)\ }\left[A+Bt\right].
\end{equation}
This case is called critical damped temperature oscillations.

\item[(iii)]  When  ${\omega }_n>{1}/{(2{\tau }_{q1})}$, we have then two distinct complex roots ${\lambda }_{1,2}=-\frac{1}{2{\tau }_{q1}}\pm \frac{\imath }{2{\tau }_{q1}}\sqrt{4{\tau }^2_{q1}{\omega }^2_n-1}$, and the solution is given as

\begin{equation} \label{eq32}
{\mathrm{\Gamma }}_n\left(t\right)={\exp \left(-\frac{t}{2{\tau }_{q1}}\right)\ }\left[A \ {\cos \left(\frac{t}{2{\tau }_{q1}}\sqrt{4{\tau }^2_{q1}{\omega }^2_n-1}\right)\ }+B \ {\sin \left(\frac{t}{2{\tau }_{q1}}\sqrt{1-4{\tau }^2_{q1}{\omega }^2_n}\right)\ }\right].
\end{equation}
This case is called underdamped temperature oscillations.
\end{enumerate}
 In \eqref{eq30}-\eqref{eq32}, the coefficients $A$ and $B$ will be determined later. The forms \eqref{eq30} and \eqref{eq32} can be converted to each other with specific choices for the coefficients. Therefore, in the classical damped wave model, the range of natural frequency covers all three cases of thermal waves: overdamped, critically damped and underdamped thermal waves. The critical frequency of the classical damped wave model is defined by

\begin{equation} \label{eq33}
{\widehat{\omega }}_{nc}=\frac{1}{2{\tau }_{q1}}=\frac{1}{2{\tau }_F}.
\end{equation}

The critical frequency ${\widehat{\omega }}_{nc}$ is clearly measured on the terahertz scale. For example, near the temperature room $T_0=300\ \mathrm{K}$, the critical frequency for classical damped wave model is ${\widehat{\omega }}_{nc}=12.5\ \mathrm{THz}$ for gold, ${\widehat{\omega }}_{nc}=16.7\ \mathrm{THz}$ for copper, and ${\widehat{\omega }}_{nc}=100\ \mathrm{THz}$ for lead, i.e., it may fall within the Terahertz gap in which receiving or generating terahertz signals is unfeasible \cite{RN43}.

\subsection{ Two-Step damped wave solution}

\noindent  In the case of considering the heat capacity of the metal lattice, i.e., ${C_l}/{G}$ is a non-negligible quantity, the characteristic equation is given through the cubic equation \eqref{eq13} and it can be simplified to the form:

\begin{equation} \label{eq34}
{\lambda }^3+\left(\frac{1}{{\tau }_{q1}}+\frac{1}{{\tau }_{q2}}\right){\lambda }^2+\left(\frac{1}{{\tau }_{q1}{\tau }_{q2}}+{\omega }^2_n\right)\lambda +\frac{{\omega }^2_n}{{\tau }_T}=0.
\end{equation}

Using DERIVE 6.1 to solve the expression $\left\{a_3{\lambda }^3+a_2{\lambda }^2+a_1\lambda +a_0=0\right\}$ for $\lambda $, where $a_3$, $a_2$, $a_1$ and $a_0$ are given real coefficients, one can obtain the following ``specific form'' of characteristic roots:

\begin{eqnarray}\label{eq35}
	\nonumber
	{\lambda }_1=\frac{2\sqrt{{\mathrm{\Delta }}_0}{\sin \left(\frac{1}{3}{{\sin}^{-1} \left(\frac{{\mathrm{\Delta }}_1\mathrm{sign}\left(a_3\right)}{2\sqrt{{\mathrm{\Delta }}^3_0}}\right)\ }\right)\ }}{3\left|a_3\right|}-\frac{a_2}{3a_3},\qquad {\lambda }_2=\frac{2\sqrt{{\mathrm{\Delta }}_0}{\cos \left(\frac{1}{3}{{\cos}^{-1} \left(-\frac{{\mathrm{\Delta }}_1\mathrm{sign}\left(a_3\right)}{2\sqrt{{\mathrm{\Delta }}^3_0}}\right)\ }\right)\ }}{3\left|a_3\right|}-\frac{a_2}{3a_3}, \\
{\lambda }_3=-\frac{2\sqrt{{\mathrm{\Delta }}_0}{\sin \left(\frac{\pi }{3}+\frac{1}{3}{{\sin}^{-1} \left(\frac{{\mathrm{\Delta }}_1\mathrm{sign}\left(a_3\right)}{2\sqrt{{\mathrm{\Delta }}^3_0}}\right)\ }\right)\ }}{3\left|a_3\right|}-\frac{a_2}{3a_3},
\end{eqnarray}
where ${\mathrm{\Delta }}_0$ and ${\mathrm{\Delta }}_1$ are respectively the discriminants of zero and one, given by

\begin{equation}\label{eq36}
{\mathrm{\Delta }}_0=a^2_2-3a_1a_3,\ \ \ {\mathrm{\Delta }}_1=2a^3_2-9a_1a_2a_3+27a_0a^2_3.
\end{equation}

It is known that the underdamped thermal wave occurs wherever it is possible to write the time-dependent amplitude in terms of cosine or sine functions. In view of the roots \eqref{eq35}, we deduce that the negativity of the discriminant ${\mathrm{\Delta }}_0$ is a sufficient condition to obtain a complex root, namely,

\begin{equation}\label{eq37}
{\mathrm{\Delta }}_0<0.
\end{equation}

One further possibility to obtain complex roots is that 

\begin{equation}\label{eq38}
{\mathrm{\Delta }}_0>0,\qquad \left|\frac{{\mathrm{\Delta }}_1}{2\sqrt{{\mathrm{\Delta }}^3_0}}\right|>1,
\end{equation}
where ${{\sin}^{-1} \left(\theta \right)\ }$ and ${{\cos}^{-1} \left(\theta \right)\ }$ are complex-valued functions whenever $\left|\theta \right|>1$, furthermore, sine and cosine functions map $\mathbb{C}$ to $\mathbb{C}$. Replacing $a_3$, $a_2$, $a_1$ and $a_0$ with the coefficients of ${\lambda }^3$, ${\lambda }^2$, $\lambda $ and the absolute term in equation \eqref{eq34}, and applying the first sufficient condition for complex roots, \eqref{eq36} and \eqref{eq37}, we obtain  

\begin{equation}\label{eq39}
{\omega }_n>{\omega }_{nc}=\sqrt{\frac{1}{3}\left(\frac{1}{{\tau }^2_{q1}}-\frac{1}{{\tau }_{q1}{\tau }_{q2}}+\frac{1}{{\tau }^2_{q2}}\right)}=\sqrt{\frac{1}{3}\left(\frac{1}{{\tau }^2_F}+\frac{G\left(C_e+C_l\right)}{C_eC_l}\left(\frac{G\left(C_e+C_l\right)}{C_eC_l}-\frac{1}{{\tau }_F}\right)\right)},
\end{equation}
where ${\omega }_{nc}$, defined in \eqref{eq39}, refers to the critical frequency in the case of two-step damped thermal wave model. Therefore, if the natural frequency exceeds this threshold, ${\omega }_n>{\omega }_{nc}$, the temperature oscillation may be of the underdamped type. The value of critical frequency provided by \eqref{eq39} generalizes the classical threshold \eqref{eq33} derived previously by Tzou \cite{RN35}. The existence of such a critical value can be guaranteed by requiring the nonnegativity of the magnitude under the square root of \eqref{eq39}. In other words, if we assume that 

\begin{equation}\label{eq40}
{\tau }_{q1}<{\tau }_{q2}\qquad \mathrm{or}\qquad{\tau }_F<\frac{C_eC_l}{G\left(C_e+C_l\right)},
\end{equation}
then we have ${\tau }^2_{q1}<{\tau }_{q1}{\tau }_{q2}$, which leads to ${1}/{{\tau }^2_{q1}}>{1}/{{\tau }_{q1}{\tau }_{q2}}$, or alternatively ${1}/{{\tau }^2_{q1}}+{1}/{{\tau }^2_{q1}}-{1}/{{\tau }_{q1}{\tau }_{q2}}>0$, therefore ${\omega }_{nc}$ derived in \eqref{eq39} is a positive real. The case ${\tau }_{q1}<{\tau }_{q2}$ or ${\tau }_F<\frac{C_eC_l}{G\left(C_e+C_l\right)}$ is verified for the sample of metals chosen in this study, i.e., gold, copper, and lead, refer to Table \ref{tabel1}.

On the other hand, the second condition that may lead to a complex root, \eqref{eq38}, adds an additional domain of frequency to \eqref{eq39}. The first condition of \eqref{eq38} leads to ${\omega }_n<{\omega }_{nc}$, while the second condition of \eqref{eq38} yields the following six order inequality

\begin{multline}\label{eq41}
{\omega }^6_n+\left[\frac{3}{4}{\left(\frac{3}{{\tau }_T}-\frac{1}{{\tau }_{q1}}-\frac{1}{{\tau }_{q2}}\right)}^2-3{\omega }^2_{nc}\right]{\omega }^4_n+ \\ 
\left[3{\omega }^4_{nc}+\frac{1}{6}\left(\frac{1}{{\tau }_{q1}}+\frac{1}{{\tau }_{q2}}\right)\left(\frac{3}{{\tau }_T}-\frac{1}{{\tau }_{q1}}-\frac{1}{{\tau }_{q2}}\right)\left(\frac{2}{{\tau }^2_{q1}}-\frac{5}{{\tau }_{q1}{\tau }_{q2}}+\frac{2}{{\tau }^2_{q2}}\right)\right]{\omega }^2_n+ \\
\frac{1}{108}{\left(\frac{1}{{\tau }_{q1}}+\frac{1}{{\tau }_{q2}}\right)}^2{\left(\frac{2}{{\tau }^2_{q1}}-\frac{5}{{\tau }_{q1}{\tau }_{q2}}+\frac{2}{{\tau }^2_{q2}}\right)}^2-{\omega }^6_{nc}>0,
\end{multline}
where ${\omega }_{nc}$ is given by \eqref{eq39}. Inequality \eqref{eq41} defines a second lower critical frequency ${\omega }_{nc0}$, but it is difficult to obtain explicit forms for the six roots of \eqref{eq41}. Therefore, the second lower limit of critical frequency ${\omega }_{nc0}$ will be appreciated numerically for Au, Cu and Pb using MATHCAD. Upon combining the first condition of \eqref{eq38} and the second condition \eqref{eq41}, we define the new domain of frequency such that for natural frequency satisfying 

\begin{equation}\label{eq42}
{\omega }_{nc0}<{\omega }_n<{\omega }_{nc},
\end{equation}
the roots of \eqref{eq34} may be complex and the temperature oscillations are then underdamped, where  ${\omega }_{nc0}$ is the positive real root of \eqref{eq41}. In Table \ref{tabel2}, the two critical frequencies ${\omega }_{nc}$ and ${\omega }_{nc0}$  from respectively \eqref{eq39} and \eqref{eq41} for three metals are computed.
	
\begin{table}
	\centering
	\caption{The critical frequency \textit{${\omega }_{nc}\ $}derived in \eqref{eq39} and the lowest critical frequency computed numerically from inequality \eqref{eq41} for Cu, Au and Pb near room temperature \textit{$T_0=300\ K$}.}
	\label{tabel2}
	\scalebox{1}{
		\begin{tabular}{||ccc||} \hline \hline 
			\text{} & ${\omega }_{nc0}$ & ${\omega }_{nc}$  \\ 
			\textbf{Metal} &  THz (${10}^{12}$  Hz) &  THz (${10}^{12}$  Hz) \\ 
			\hline \hline 
			\textbf{Cu} & 15.52 & 18.62 \\ \hline 
			\textbf{Au } & 11.83 & 14.06 \\ \hline 
			\textbf{Pb} & 97.04 & 113.78 \\ 
			\hline \hline 
		\end{tabular}
	}
\end{table}

From the above analysis, we have two adjacent frequency intervals which lead to complex roots for equation \eqref{eq34}, and thus evidencing to the possibility of underdamped temperature oscillations when the natural frequency falls within them. The first domain is given by equation \eqref{eq39}, ${\omega }_n>{\omega }_{nc}$, and the second domain is given by \eqref{eq42}, ${\omega }_{nc0}<{\omega }_n<{\omega }_{nc}$. Otherwise, the thermal waves are predicted to be overdamped. In order to check these mathematical conjectures, we must invoke to illuminating numerical examples. Before moving on, we suggest the solution of \eqref{eq34} on the form

\begin{equation}\label{eq43}
{\mathrm{\Gamma }}_n\left(t\right)=A \ {\exp \left({\lambda }_1t\right)\ }+B \ {\exp \left({\lambda }_2t\right)\ }+C \ {\exp \left({\lambda }_3t\right)\ }, 
\end{equation}
where ${\lambda }_1$, ${\lambda }_2$ and ${\lambda }_3$ are given by \eqref{eq35} and $A$, $B$ and $C$ are unknown coefficients, definitely different from the ones used in equations \eqref{eq30}-\eqref{eq32}, will be determined later. Rewriting equation \eqref{eq35} as $\left(\lambda +{\lambda }_1\right)\left(\lambda +{\lambda }_2\right)\left(\lambda +{\lambda }_3\right)=0$, it yields that ${\lambda }_1+{\lambda }_2+{\lambda }_3={1}/{{\tau }_{q1}}+{1}/{{\tau }_{q2}}$, i.e., the sum of the roots \eqref{eq35} must be real, which means that either the three roots are real, or one is real and the other two roots are complex conjugates. If we suppose that ${\lambda }_1$ is the real root in the case of underdamping, then ${\lambda }_2$ and ${\lambda }_3$ are given as ${\lambda }_{2,3}=r_0\pm \imath I_0$, and the time-dependent amplitude can be written as

\begin{equation}\label{eq44}
{\mathrm{\Gamma }}_n\left(t\right)=A \ {\exp \left({\lambda }_1t\right)\ }+{\exp \left(r_0t\right)\ }\left[B \ {\cos \left(I_0t\right)\ }+C \ {\sin \left(I_0t\right)\ }\right],
\end{equation}
and again both forms \eqref{eq43} and \eqref{eq44} can be converted by specific choices for $A$, $B$ and $C$. 

The question arisen now is that what is the condition leading to the critical damping in the case of two-step damped wave model? By the critical damping here, we mean that at least two roots of ${\lambda }_1$, ${\lambda }_2$ and ${\lambda }_3$ are equal. The conditions of three equal roots which lead to the solution ${\mathrm{\Gamma }}_n\left(t\right)={\exp \left(\lambda t\right)\ }\left(A+Bt+Bt^2\right)$ is not satisfied in our case. Indeed, we have that there are such equal roots for the cubic equation if ${\delta }_1={\delta }_2={\delta }_3=0$ where ${\delta }_1$, ${\delta }_2$ and ${\delta }_3$ are given by \cite{RN44}

\begin{equation}\label{eq45}
{\delta }_1=\frac{1}{3}a_1a_3-\frac{1}{9}a^2_2,\ \ \ {\delta }_2=\frac{1}{3}a_0a_3-\frac{1}{9}a_1a_2,\ \ \ {\delta }_3=\frac{1}{3}a_0a_2-\frac{1}{9}a^2_1.
\end{equation}

Utilizing the above quantities in \eqref{eq45} with the coefficients of the cubic characteristic equation \eqref{eq34}, the first condition ${\delta }_1=0$ leads to ${\omega }_n={\omega }_{nc}$, while the second condition ${\delta }_2=0$ gives a complex value for ${\omega }_n$, which is rejected and then ${\delta }_2$ could not vanish, i.e., there are no a critical damping case in which the three roots are equal. The second possibility of critical damping, in which there are two roots are equal real numbers, occurs when  ${\delta }^2_2=4{\delta }_1{\delta }_3$, \cite{RN44}, which reduces to \eqref{eq41}, namely it is equivalent to ${\omega }_n={\omega }_{nc0}$. Therefore, the only critical damping occurs when the natural frequency equals the lowest critical frequency, and the solution then can be written as

\begin{equation}\label{eq46}
{\mathrm{\Gamma }}_n\left(t\right)=A{\exp \left({\lambda }_1t\right)\ }+{\exp \left({\lambda }_2t\right)\ }\left[B+Ct\right].
\end{equation}

In the end of this section, we have three possible forms of the solution of \eqref{eq25}. The first, given by \eqref{eq43}, works for the natural frequency ${\omega }_n<{\omega }_{nc0}$, wherein the characteristic equation \eqref{eq34} has three distinct real roots. The second given by \eqref{eq44}, is valid when the natural frequency exceeds ${\omega }_{nc0}$ wherein the characteristic equation \eqref{eq34} has one real root and two complex conjugates. The third one given by \eqref{eq46} occurs when ${\omega }_n={\omega }_{nc0}$, i.e., the case of critical damping.

\subsection{ Illuminating examples}

To check whether the two frequency domains \eqref{eq39} and \eqref{eq42} correspond to underdamped thermal waves or not, we invoke practical numerical example on Gold (Au). Let us consider the following initial conditions for \eqref{eq30}, \eqref{eq31} or \eqref{eq32}

\begin{equation}\label{eq47}
	T\left(x,0\right)={\theta }_0\left(x\right),\ \ \ {\left.\frac{\partial T\left(x,t\right)}{\partial t}\right|}_{t=0}={\theta }_1\left(x\right),
\end{equation}
and the following initial conditions for \eqref{eq43}, \eqref{eq44} or \eqref{eq46}

\begin{equation}\label{eq48}
	T\left(x,0\right)={\theta }_0\left(x\right),\ \ \ {\left.\frac{\partial T\left(x,t\right)}{\partial t}\right|}_{t=0}={\theta }_1\left(x\right),\ \ \ {\left.\frac{{\partial }^2T\left(x,t\right)}{\partial t^2}\right|}_{t=0}={\theta }_2\left(x\right),
\end{equation}
where ${\theta }_0\left(x\right)$, ${\theta }_1\left(x\right)$ and ${\theta }_2\left(x\right)$ are given functions. The initial conditions \eqref{eq47} and \eqref{eq48} together with the representation \eqref{eq24} yield

\begin{equation}\label{eq49}
	{\mathrm{\Gamma }}_n\left(0\right)={\theta }_{n0},\ \ \ {\left.\frac{\partial {\mathrm{\Gamma }}_n\left(t\right)}{\partial t}\right|}_{t=0}={\theta }_{n1},\ \ \ {\left.\frac{{\partial }^2{\mathrm{\Gamma }}_n\left(t\right)}{\partial t^2}\right|}_{t=0}={\theta }_{n2},
\end{equation}
where 

\begin{equation}\label{eq50}
	{\theta }_{ni}=\frac{\int^L_0{{\theta }_i\left(x\right){\phi }_n\left(x\right)}\mathrm{d}x}{\int^L_0{{\phi }^2_n\left(x\right)}\mathrm{d}x}, \qquad i=0,1,2.
\end{equation}

The first two initial conditions of \eqref{eq49} with the solution \eqref{eq30} determine the coefficients of time-dependent amplitude ${\mathrm{\Gamma }}_n\left(t\right)$ for the temperature governed by the classical damped wave model:

\begin{eqnarray}\label{eq51}
	\nonumber
	A=\frac{1}{2}\left[{\theta }_{n0}\left(1+\frac{1}{\sqrt{1-4{\tau }^2_{q1}{\omega }^2_n}}\right)+\frac{2{\tau }_{q1}{\theta }_{n1}}{\sqrt{1-4{\tau }^2_{q1}{\omega }^2_n}}\right], \\
	B=\frac{1}{2}\left[{\theta }_{n0}\left(1-\frac{1}{\sqrt{1-4{\tau }^2_{q1}{\omega }^2_n}}\right)-\frac{2{\tau }_{q1}{\theta }_{n1}}{\sqrt{1-4{\tau }^2_{q1}{\omega }^2_n}}\right],
\end{eqnarray}
while the three initial conditions of \eqref{eq49} with \eqref{eq43} yield the coefficients of time-dependent amplitude ${\mathrm{\Gamma }}_n\left(t\right)$ for the temperature governed by the two-step damped wave model:

\begin{eqnarray}\label{eq52}
	\nonumber
	A=\frac{{\theta }_{n2}-\left({\lambda }_2+{\lambda }_3\right){\theta }_{n1}+{\lambda }_2{\lambda }_3{\theta }_{n0}}{\left({\lambda }_1-{\lambda }_2\right)\left({\lambda }_1-{\lambda }_3\right)},\ \ \ B=\frac{{\theta }_{n2}-\left({\lambda }_1+{\lambda }_3\right){\theta }_{n1}+{\lambda }_1{\lambda }_3{\theta }_{n0}}{\left({\lambda }_1-{\lambda }_2\right)\left({\lambda }_3-{\lambda }_2\right)}, \\
	C=\frac{{\theta }_{n2}-\left({\lambda }_1+{\lambda }_2\right){\theta }_{n1}+{\lambda }_1{\lambda }_2{\theta }_{n0}}{\left({\lambda }_1-{\lambda }_3\right)\left({\lambda }_2-{\lambda }_3\right)}.
\end{eqnarray}

In order to capture the possible free oscillations of the temperature time-dependent amplitude ${\mathrm{\Gamma }}_n\left(t\right)$, we choose the following initial data:

\begin{equation}\label{eq53}
	{\theta }_0\left(x\right)={\theta }_0={10}^{-15}\ \mathrm{K},\qquad{\theta }_1\left(x\right)={\theta }_1=1\ \mathrm{K}{\mathrm{s}}^{-1},\qquad {\theta }_2\left(x\right)={\theta }_2=0\ \mathrm{K}{\mathrm{s}}^{-2},
\end{equation}
where we impose a very small change on the temperature at $t=0$, i.e., $\check{T}\left(x,0\right)=T_0+{10}^{-15}$, $T_0=300\ \mathrm{K}$, an initial temperature rate equals $10$, and a zero second temperature rate. Further, we choose the thickness of the plate as $L=1\ \mathrm{\mu }\mathrm{m}={10}^{-6}\ \mathrm{m}$. Inserting the initial values \eqref{eq53} into \eqref{eq50}, we obtain

\begin{equation}\label{eq54}
	{\theta }_{ni}=\frac{2v_3{\theta }_i}{{\omega }_nL}\frac{1-{\cos \left(\frac{{\omega }_nL}{v_3}\right)\ }}{1-\frac{v_3}{2{\omega }_nL}{\sin \left(\frac{2{\omega }_nL}{v_3}\right)\ }},\ \ \ \ i=0,1,2.
\end{equation}

\begin{figure} 
		\centering
\includegraphics*[scale=0.6,angle=0]{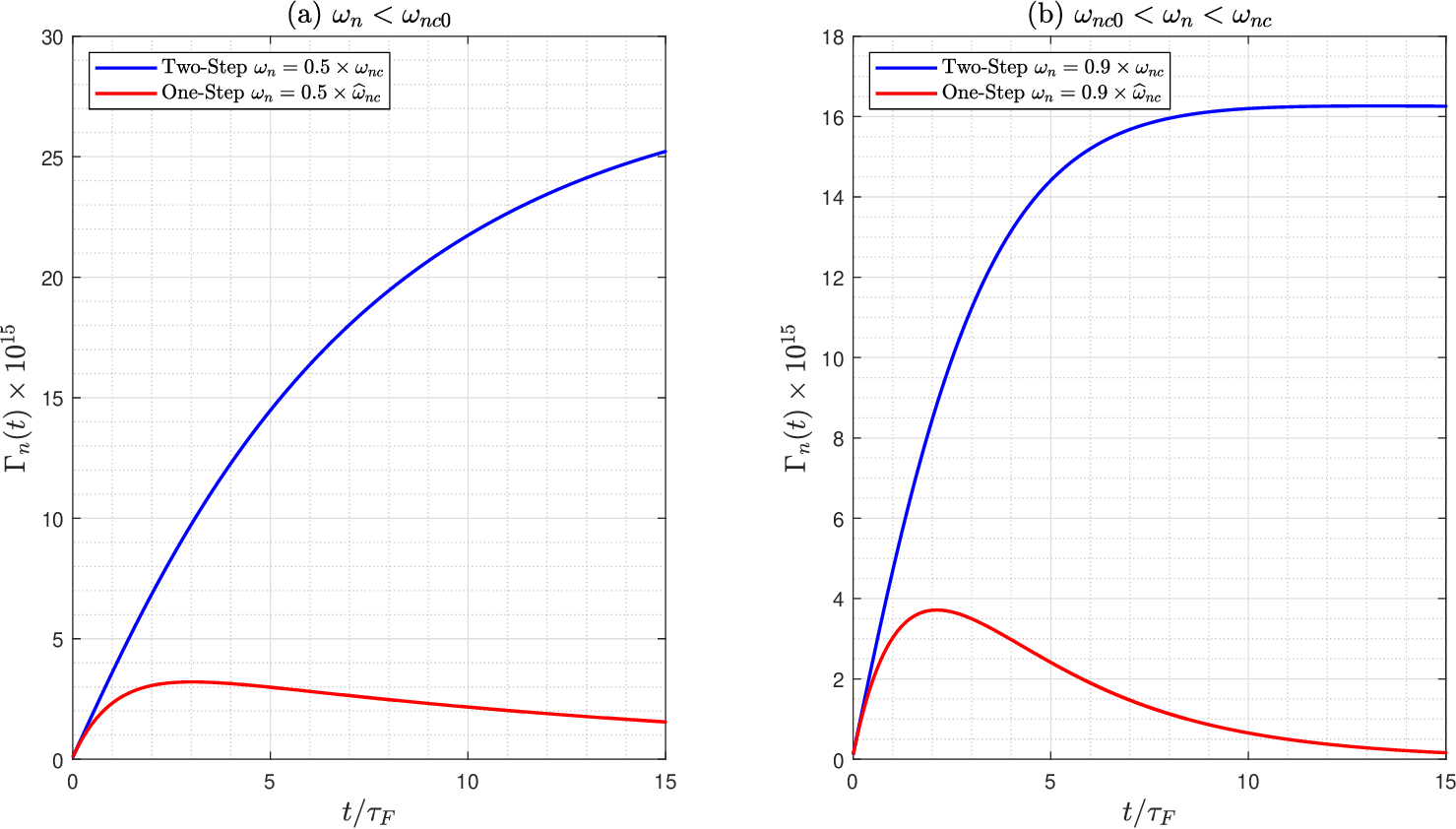} 
\includegraphics*[scale=0.6,angle=0]{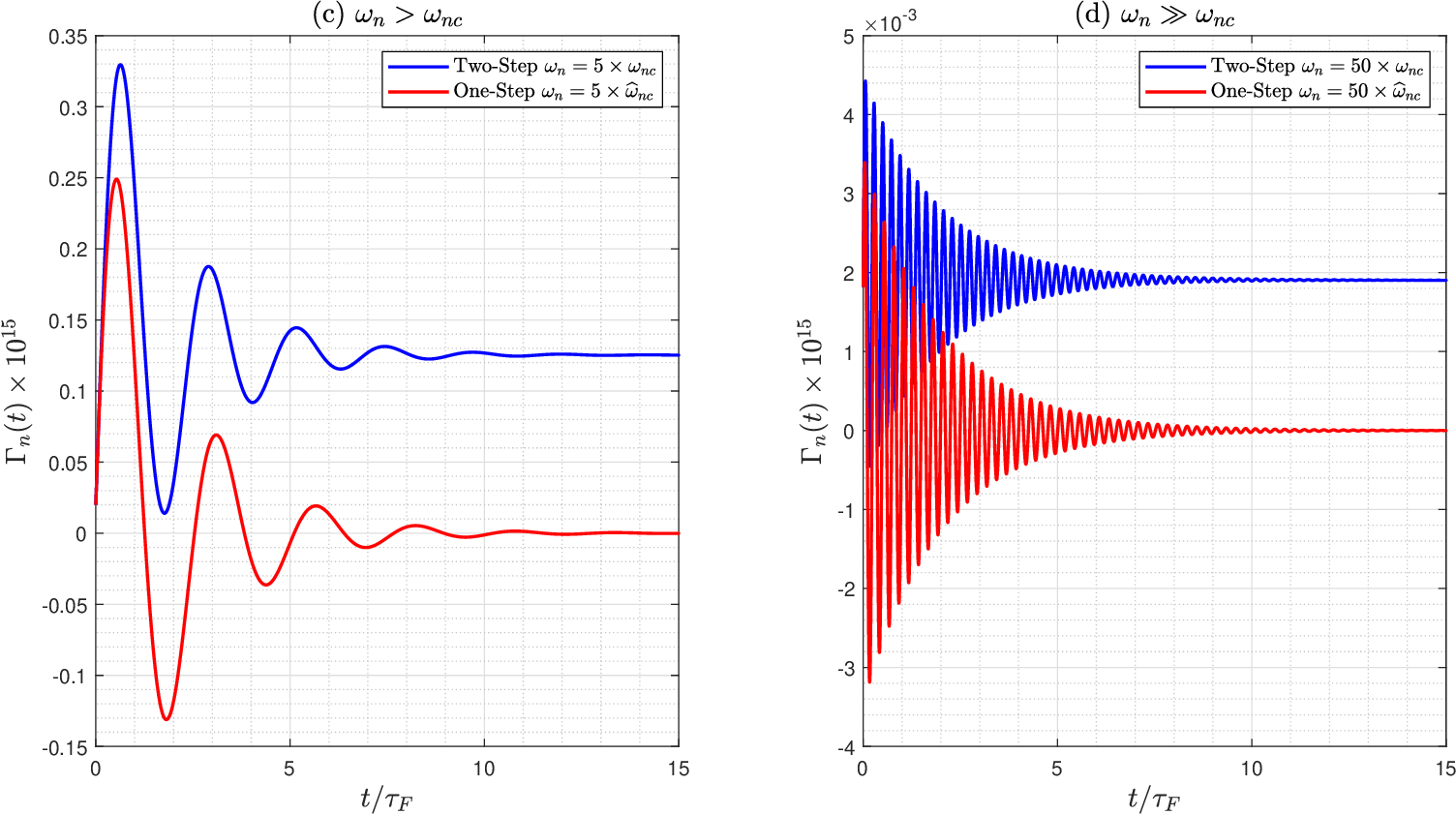} 
	\caption{Time-dependent amplitude \textit{${\mathit{\Gamma}}_n\left(t\right)$} the two-step (blue solid line) and the one-step (red solid line) models at certain values of the natural frequency. In (a) \textit{${\omega }_n=0.5 \ {\omega }_{nc}<{\omega }_{nc0}$}, (b) \textit{${\omega }_{nc0}<{\omega }_n=0.9 \ {\omega }_{nc}<{\omega }_{nc}$}, (c) \textit{${\omega }_n=5 \ {\omega }_{nc}>{\omega }_{nc}$}, (d) \textit{${\omega }_n=50 \ {\omega }_{nc}\gg {\omega }_{nc}$}.}
\label{fig1}
\end{figure}

In Fig. \ref{fig1}, The time-dependent amplitude of the one-step model \eqref{eq30} and the two-step model \eqref{eq43} with their coefficients determined in \eqref{eq51} and (\ref{eq52}) respectively, are represented at different values of the natural frequency ${\omega }_n$. The initial data \eqref{eq54} are natural frequency dependent. We focus attention on three frequency domains: (a) ${\omega }_n<{\omega }_{nc0}$, (b)\textit{ }${\omega }_{nc0}<{\omega }_n<{\omega }_{nc}$, and (c)\textit{ }${\omega }_n\mathrm{>}{\omega }_{nc}$. Therefore, the crossover from overdamping (${\omega }_n<{\omega }_{nc0}$) to the underdamping ($\omega_{n}>{\omega }_{nc}$) passing over the critical damping (${\omega }_n={\omega }_{nc0}$) is shown. Although the characteristic roots corresponding to the frequency domain ${\omega }_{nc0}<{\omega }_n<{\omega }_{nc}$ are complex numbers, the oscillations are not observed within this domain, see Fig. \ref{fig1} (b). In Table \ref{tabel3}, we calculate the characteristic roots corresponding to certain values of the natural frequency. 

\begin{table}
	\centering
	\caption{The characteristic roots \eqref{eq35} for Gold (Au) near room temperature \textit{$T_0=300\ K$} at different natural frequency values..}
	\label{tabel3}
	\scalebox{0.81}{
	\begin{tabular}{||cccccc||} \hline \hline
		&$\omega_n=0.5 \ \omega_{nc}<\omega_{nc0}$&$\omega_n=\omega_{nc0}$&$\omega_{nc0}<\omega_n<\omega_{nc}$&$\omega_n=\omega_{nc}$&$\omega_n=5 \ \omega_{nc}>\omega_{nc}$ \\ 
		\textbf{Roots}&$\left( 7.031 \ \mathrm{THz}\right) $&$\left( 11.83 \ \mathrm{THz}\right) $&$\left( 12.66 \ \mathrm{THz}\right) $&$\left( 14.06 \ \mathrm{THz}\right) $&$\left( 70.31 \ \mathrm{THz}\right) $ \\
		\hline \hline 
		\textbf{${\boldsymbol{\lambda }}_{\boldsymbol{1}}$} & $-3.65\times {10}^{12}$ & $-1.32\times {10}^{13}$ & $-1.32\times {10}^{13}+4.49\ \imath \times {10}^{12}$ & $-1.32\times {10}^{13}+7.6\ \imath \times {10}^{12}$ & $-1.11\times {10}^{10}$ \\ \hline 
		\textbf{${\boldsymbol{\lambda }}_{\boldsymbol{2}}$} & $-6.68\times {10}^9$ & $-9.04\times {10}^9$ & $-9.27\times {10}^9$ & $-9.58\times {10}^9$ & $-1.32\times {10}^{13}+6.93\ \imath \times {10}^{13}$ \\ \hline 
		\textbf{${\boldsymbol{\lambda }}_{\boldsymbol{3}}$} & $-2.27\times {10}^{13}$ & $-1.32\times {10}^{13}$ & $-1.32\times {10}^{13}-4.49\ \imath \times {10}^{12}$ & $-1.32\times {10}^{13}-7.6\ \imath \times {10}^{12}$ & $-1.32\times {10}^{13}-6.93\ \imath \times {10}^{13}$ \\ 
		\hline \hline
	\end{tabular}
}
\end{table}

We note that two complex roots are always resulted within the domain ${\omega }_{nc0}\mathrm{<}{\omega }_n\mathrm{<}{\omega }_{nc}$. These complex roots $r_0\pm \imath I_0$ are the reason behind existing damped wave solution on the form ${\exp \left(r_0t\right)\ }\left[B{\cos \left(I_0t\right)\ }+C{\sin \left(I_0t\right)\ }\right]$, refer to equation \eqref{eq44}. Because damping exponent is greater than the wave part, the damping prevails within this frequency domain and oscillations become not clear as in Fig. \ref{fig1} (b). When the natural frequency becomes large enough, ${\omega }_n\mathrm{>}{\omega }_{nc}$, the damping exponent $r_0$ is on the same order as the wave part, the oscillations can be observed.

\begin{figure} 
		\centering
\includegraphics*[scale=0.6,angle=0]{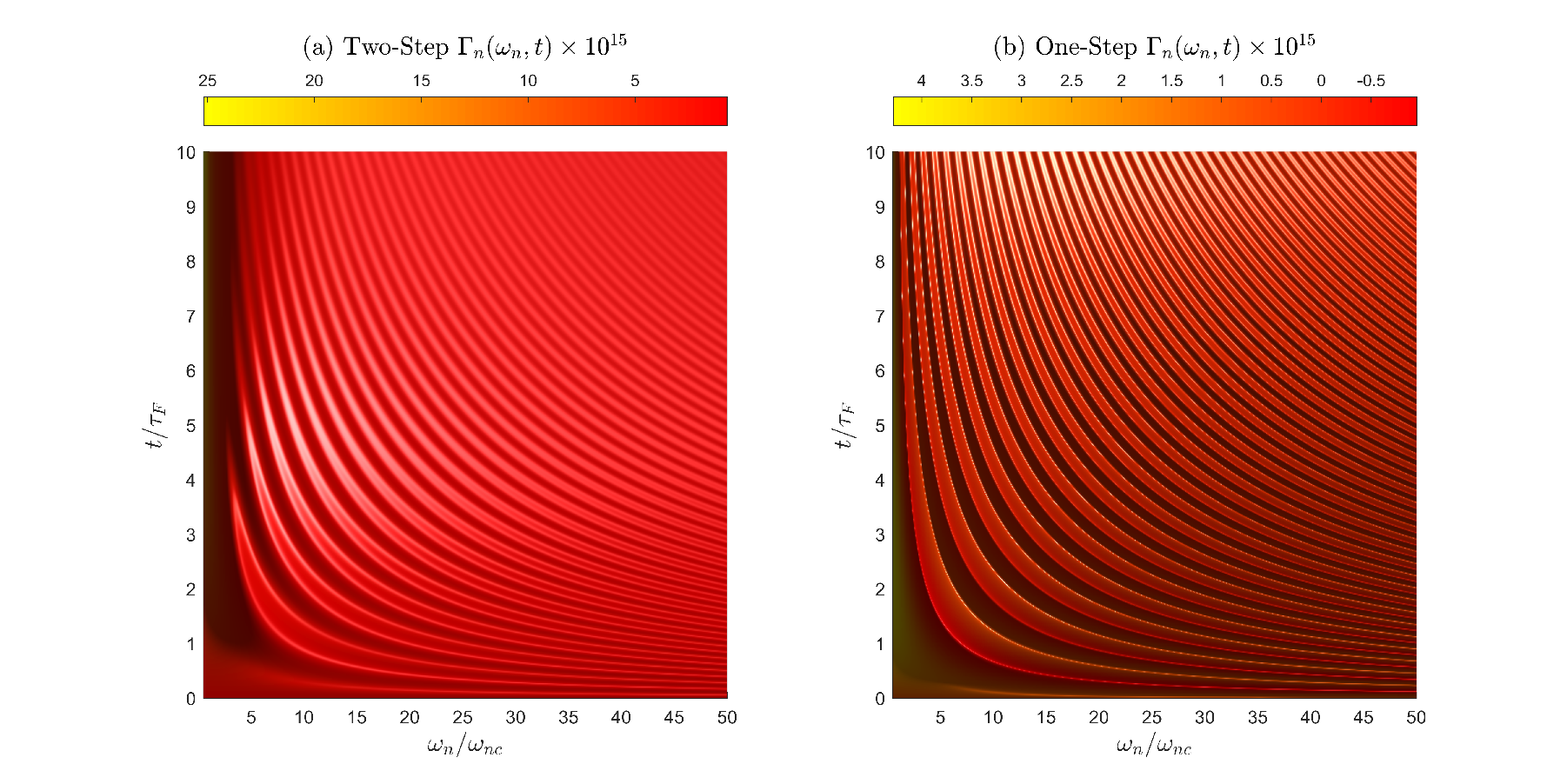} 
\caption{ The time-dependent temperature amplitude as a function of time \textit{$t$} and the natural frequency \textit{${\omega }_n$} for both (a) Two-Step model, (b) One-Step model under the effect of initial temperature rate and small initial temperature. The initial data are fixed at \textit{${\theta }_{n0}={10}^{-15}$}, \textit{${\theta }_{n1}=0.1$} and \textit{${\theta }_{n2}=0$}.}
\label{fig2}
\end{figure}

\begin{figure}
		\centering
\includegraphics*[scale=0.6,angle=0]{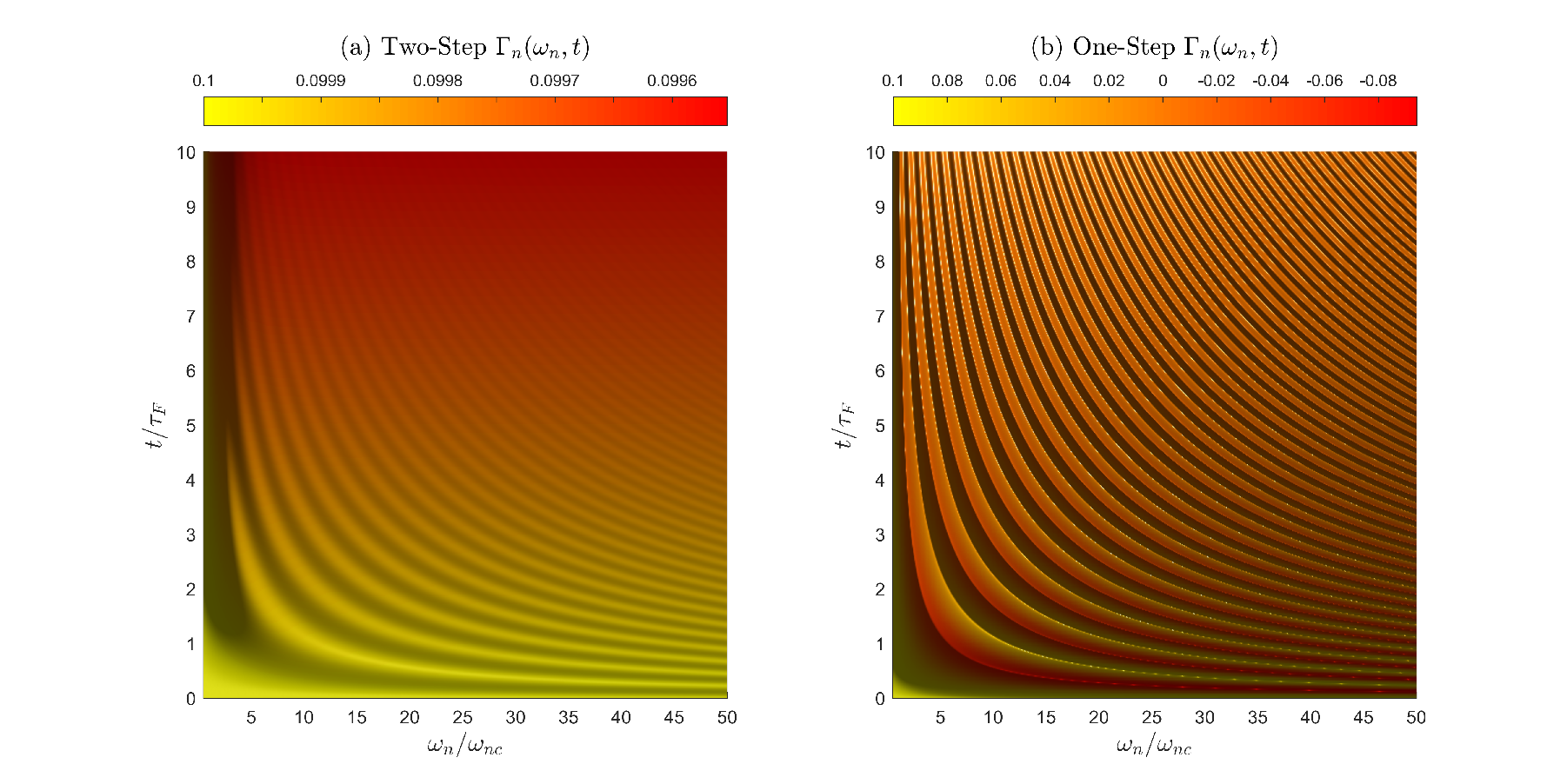} 
\caption{ The time-dependent temperature amplitude as a function of time \textit{$t$} and the natural frequency \textit{${\omega }_n$} for both (a) Two-Step model; (b) One-Step model under the effect of initial temperature only. The initial data are fixed at \textit{${\theta }_{n0}=0.1$}, \textit{${\theta }_{n1}=0$} and \textit{${\theta }_{n2}=0$}. }
\label{fig3}
\end{figure}

We find that during this small variation of initial temperature, ${\theta }_0={10}^{-15}$, the temperature oscillations are easily visible in the two-step model. On the contrary, the temperature oscillations of the classical damped wave model (or one-step model) are clear for any initial temperature variations, e.g., ${\theta }_0=10$. On the other hand, the initial temperature rate increases the visibility of the temperature oscillations through increasing their amplitudes. Increasing the initial temperature rate causes the convergence between the amplitudes of both two-step and one-step temperatures. In order to completely depict the effect of frequency domain on the temperature oscillations, we draw the time-dependent amplitude as a function of the time and the natural frequency, ${\mathrm{\Gamma }}_n\left({\omega }_n,t\right)$ in Fig. \ref{fig2} for the two-step and the one-step models. To avoid the effect of frequency-dependent initial data \eqref{eq54}, we fix them at mean values: ${\theta }_{n0}\mathrm{=}{\mathrm{10}}^{\mathrm{-}\mathrm{15}}$\textit{, }${\theta }_{n\mathrm{1}}\mathrm{=0.1}$\textit{ }and\textit{ }${\theta }_{n\mathrm{2}}\mathrm{=0}$. It is clear that the temperature amplitude oscillates after exceeding a certain frequency threshold in both models, ${\omega }_{nc}$ for the two-step model and ${\widehat{\omega }}_{nc}$ for the one-step model. The effect of lonely initial temperature on the oscillations of time-dependent amplitude is shown in Fig. \ref{fig3}. With the fixed values ${\theta }_{n0}\mathrm{=0.1}$\textit{, }${\theta }_{n\mathrm{1}}\mathrm{=0}$\textit{ }and\textit{ }${\theta }_{n\mathrm{2}}\mathrm{=0}$, we see that the non-zero initial temperature causes small oscillation which may be barely observed in the 2D plot, note its variations in the colorbar of Fig. \ref{fig3} (a) and (b) wherein the change of two-step amplitude lies within ${\mathrm{\Gamma }}_n\left({\omega }_n,t\right)\in \left(0.09996,0.1\right)$ and the change of the one-step amplitude lies within ${\mathrm{\Gamma }}_n\left({\omega }_n,t\right)\in \left(-0.08,0.1\right)$.

\section{ Thermal resonance phenomenon}
\label{sec4}
In this section, we present a mathematical conjecture to what-so-called thermal resonance phenomenon in which the temperature amplitude is amplified under a certain value of the external applied frequency. Unlike the classical damped wave (hyperbolic one-step) model, the hyperbolic two-step model distinguishes the electron temperature and the lattice temperature \eqref{eq10}. Both thermal waves of the electron and lattice temperature propagate with the same speed $v_3$ and their oscillations are guaranteed to be underdamped whenever the natural frequency of the thermal system exceeds a critical value \eqref{eq39}. Let the external heat source $Q\left(x,t\right)$ oscillate on a frequency $\mathrm{\Omega }$, and be determined by 

\begin{equation}\label{eq55}
Q\left(x,t\right)=Q_0g\left(x\right){\exp \left(\imath \mathrm{\Omega }t\right)\ },
\end{equation}
where $Q_0$ and $g\left(x\right)$ are respectively the strength and the spatial distribution of the heat source. The external source frequency refers to the number of thermal pulses per second. Therefore, we can appreciate the single thermal pulse duration as $t_p=1/\mathrm{\Omega }$, that defines the nature of thermalization process. For example, if $\mathrm{\Omega }\mathrm{=1\ THz}$, then the thermal pulse duration is about 1 picosecond, i.e., the heating process is ultrashort.  Using the external heat source \eqref{eq55} into equation \eqref{eq10} we get

\begin{multline}\label{eq56}
\left\{\frac{1}{\alpha }\frac{\partial }{\partial t}+\left(\frac{1}{v^2_1}+\frac{1}{v^2_2}\right)\frac{{\partial }^2}{\partial t^2}+\frac{{\tau }_T}{v^2_3}\frac{{\partial }^3}{\partial t^3}\right\}\left[ \begin{array}{c}
T_e \\ 
T_l \end{array}
\right]=\\
\left(1+{\tau }_T\frac{\partial }{\partial t}\right)\frac{{\partial }^2}{\partial x^2}\left[ \begin{array}{c}
T_e \\ 
T_l \end{array}
\right]+\frac{Q_0}{k_e}g\left(x\right)\left[ \begin{array}{c}
\left(1+\imath \mathrm{\Omega }{\tau }_{q1}\right)\left(1+\imath \mathrm{\Omega }{\tau }_T\right) \\ 
\left(1+\imath \mathrm{\Omega }{\tau }_{q1}\right) \end{array}
\right]{\exp \left(\imath \mathrm{\Omega }t\right)\ }.
\end{multline}

Introducing the following modal representation:

\begin{equation}\label{eq57}
\left[ \begin{array}{c}
T_e \\ 
T_l \end{array}
\right]=\sum^{\infty }_{n=1}{\left[ \begin{array}{c}
{\mathrm{\Gamma }}_{ne}\left(t\right) \\ 
{\mathrm{\Gamma }}_{nl}\left(t\right) \end{array}
\right]{\phi }_n\left(x\right)},
\end{equation}
where ${\phi }_n\left(x\right)$ is the orthogonal set \eqref{eq23}, and ${\mathrm{\Gamma }}_{ne}\left(t\right)$ and ${\mathrm{\Gamma }}_{nl}\left(t\right)$ are the time-dependent amplitudes of the electron and lattice temperature respectively, equation \eqref{eq56} can be simplified to

\begin{multline}\label{eq58}
\left\{\frac{{\mathrm{d}}^3}{\mathrm{d}t^3}+\left(\frac{1}{{\tau }_{q1}}+\frac{1}{{\tau }_{q2}}\right)\frac{{\mathrm{d}}^2}{\mathrm{d}t^2}+\left(\frac{1}{{\tau }_{q1}{\tau }_{q2}}+{\omega }^2_n\right)\frac{\mathrm{d}}{\mathrm{d}t}+\frac{{\omega }^2_n}{{\tau }_T}\right\}\left[ \begin{array}{c}
{\mathrm{\Gamma }}_{ne}\left(t\right) \\ 
{\mathrm{\Gamma }}_{nl}\left(t\right) \end{array}
\right]=\\
\frac{Q_0v^2_3g_n}{k_e{\tau }_T}\left[ \begin{array}{c}
\left(1+\imath \mathrm{\Omega }{\tau }_{q1}\right)\left(1+\imath \mathrm{\Omega }{\tau }_T\right) \\ 
\left(1+\imath \mathrm{\Omega }{\tau }_{q1}\right) \end{array}
\right]{\exp \left(\imath \mathrm{\Omega }t\right)\ },
\end{multline}
where the relations \eqref{eq11}, \eqref{eq21} and the orthogonality of $\left\{{\phi }_n\left(x\right),\ n=1,2,\dots \right\}$ have been utilized during the simplifications, and $g_n$ is given by

\begin{equation}\label{eq59}
g_n=\frac{\int^L_0{g\left(x\right){\phi }_n\left(x\right)}\mathrm{d}x}{\int^L_0{{\phi }^2_n\left(x\right)}\mathrm{d}x}.
\end{equation}

In response to the harmonic external excitation ${\exp \left(\imath \mathrm{\Omega }t\right)\ }$, one of the admissible forms of the time-dependent amplitude of the electron and the lattice temperatures can be written as

\begin{equation}\label{eq60}
\left[ \begin{array}{c}
{\mathrm{\Gamma }}_{ne}\left(t\right) \\ 
{\mathrm{\Gamma }}_{nl}\left(t\right) \end{array}
\right]=\left[ \begin{array}{c}
A_{ne} \\ 
A_{nl} \end{array}
\right]{\exp \left(\imath \mathrm{\Omega }t\right)\ },
\end{equation}
with the amplitudes $A_{ne}$ and $A_{nl}$. Using the admissible process into the governing equation \eqref{eq58} yields

\begin{equation}\label{eq61}
\left[ \begin{array}{c}
A_{ne} \\ 
A_{nl} \end{array}
\right]=\frac{Q_0v^2_3g_n}{k_e}\left[ \begin{array}{c}
A_{\mathrm{\Omega }e}\left(\mathrm{\Omega }\right) \\ 
A_{\mathrm{\Omega }l}\left(\mathrm{\ }\mathrm{\Omega }\right) \end{array}
\right],
\end{equation}
with $A_{\mathrm{\Omega }e}\left(\mathrm{\Omega }\right)$ and $A_{\mathrm{\Omega }l}\left(\mathrm{\ }\mathrm{\Omega }\right)$ are given by

\begin{equation}\label{eq62}
A_{\mathrm{\Omega }e}\left(\mathrm{\Omega }\right)=\frac{\left(1+\imath \mathrm{\Omega }{\tau }_{q1}\right)\left(1+\imath \mathrm{\Omega }{\tau }_T\right)}{\left[{\omega }^2_n-{\tau }_T\left(\frac{1}{{\tau }_{q1}}+\frac{1}{{\tau }_{q2}}\right){\mathrm{\Omega }}^{\mathrm{2}}\right]+\imath \mathrm{\Omega }{\tau }_T\left(\frac{1}{{\tau }_{q1}{\tau }_{q2}}+{\omega }^2_n-{\mathrm{\Omega }}^{\mathrm{2}}\right)},
\end{equation}
 \begin{equation}\label{eq63}
A_{\mathrm{\Omega }l}\left(\mathrm{\Omega }\right)=\frac{1+\imath \mathrm{\Omega }{\tau }_{q1}}{\left[{\omega }^2_n-{\tau }_T\left(\frac{1}{{\tau }_{q1}}+\frac{1}{{\tau }_{q2}}\right){\mathrm{\Omega }}^{\mathrm{2}}\right]+\imath \mathrm{\Omega }{\tau }_T\left(\frac{1}{{\tau }_{q1}{\tau }_{q2}}+{\omega }^2_n-{\mathrm{\Omega }}^{\mathrm{2}}\right)}.
\end{equation}

We note that when ${\tau }_T\to 0$ and ${{\tau }_T}/{{\tau }_{q2}}\to 1$, both frequency-dependent amplitudes \eqref{eq62} and \eqref{eq63} reduce to the classical form of the one-step model \cite{RN35}

 \begin{equation}\label{eq64}
A_{\mathrm{\Omega }}\left(\mathrm{\Omega }\right)=\frac{1+\imath \mathrm{\Omega }{\tau }_{q1}}{\left({\omega }^2_n-{\mathrm{\Omega }}^{\mathrm{2}}\right)+\frac{\imath \mathrm{\Omega }}{{\tau }_{q1}}}.
\end{equation}

Because the thermal resonance of $T_e\left(x,t\right)$ and $T_l\left(x,t\right)$ occurs when their time-dependent amplitudes \eqref{eq60} become excessive at a certain value of the excitation frequency $\mathrm{\Omega }$, then we can assume that the thermal resonance for both temperatures occurs when $\mathrm{\Omega }={\check{\mathrm{\Omega }}}_e$ and $\mathrm{\Omega }={\check{\mathrm{\Omega }}}_l$ at which both amplitudes \eqref{eq62} and \eqref{eq63} are Maximas. The critical values of external frequency can be determined by solving the following matrix equation: 

 \begin{equation}\label{eq65}
\frac{\mathrm{d}}{\mathrm{d}\mathrm{\Omega }}\left[ \begin{array}{c}
{\left|A_{\mathrm{\Omega }e}\left(\mathrm{\Omega }\right)\right|}^2 \\ 
{\left|A_{\mathrm{\Omega }l}\left(\mathrm{\Omega }\right)\right|}^2 \end{array}
\right]=\left[ \begin{array}{c}
0 \\ 
0 \end{array}
\right].
\end{equation}

The first equation of \eqref{eq65} with \eqref{eq62} result in the following eighth order (biquartic) equation 

 \begin{equation}\label{eq66}
b_4{\mathrm{\Omega }}^8+b_3{\mathrm{\Omega }}^6+b_2{\mathrm{\Omega }}^4+b_1{\mathrm{\Omega }}^2+b_0=0,
\end{equation}
where the coefficients $b_4$, $b_3$, $b_2$, $b_1$ and $b_0$ are given by

 \begin{equation}\label{eq67}
 	\begin{gathered}
 	b_4={\tau }^2_{q1}{\tau }^4_T,\qquad\ b_3=2{\tau }^2_T\left({\tau }^2_{q1}+{\tau }^2_T\right),\\
 	b_2=3{\tau }^2_T+\left({\tau }^2_{q1}+{\tau }^2_T\right)\left[{\left(\frac{{\tau }_T}{{\tau }_{q1}}+\frac{{\tau }_T}{{\tau }_{q2}}\right)}^2-2{\tau }^2_T\left(\frac{1}{{\tau }_{q1}{\tau }_{q2}}+{\omega }^2_n\right)\right]-{\tau }^2_{q1}{\tau }^2_T\left[{\tau }^2_T{\left(\frac{1}{{\tau }_{q1}{\tau }_{q2}}+{\omega }^2_n\right)}^2-2\left(\frac{{\tau }_T}{{\tau }_{q1}}+\frac{{\tau }_T}{{\tau }_{q2}}\right){\omega }^2_n\right], \\
 	b_1=2\left[{\left(\frac{{\tau }_T}{{\tau }_{q1}}+\frac{{\tau }_T}{{\tau }_{q2}}\right)}^2-2{\tau }^2_T\left(\frac{1}{{\tau }_{q1}{\tau }_{q2}}+{\omega }^2_n\right)\right]-2{\tau }^2_{q1}{\tau }^2_T{\omega }^4_n, \\
b_0={\tau }^2_T{\left(\frac{1}{{\tau }_{q1}{\tau }_{q2}}+{\omega }^2_n\right)}^2-2\left(\frac{{\tau }_T}{{\tau }_{q1}}+\frac{{\tau }_T}{{\tau }_{q2}}\right){\omega }^2_n-\left({\tau }^2_{q1}+{\tau }^2_T\right){\omega }^4_n.
\end{gathered}
\end{equation}

The biquartic characteristic equation \eqref{eq66} has eight roots for $\mathrm{\Omega }$, such that the positive real roots are the possible values of external frequency at which the amplitude $\left|A_{\mathrm{\Omega }l}\left(\mathrm{\Omega }\right)\right|$ may be maxima. It is hard to find specific expressions for those positive roots, so we have to examine \eqref{eq66} with its coefficients \eqref{eq67} numerically using a suitable symbolic program (MATHCAD) to find the critical value of natural frequency ${\omega }_n$ which leads to the existence of such a positive real root. First, we define the discriminants of zero and one for the quartic equation $b_4{\mathrm{\Omega }}^4+b_3{\mathrm{\Omega }}^3+b_2{\mathrm{\Omega }}^2+b_1\mathrm{\Omega }+b_0=0$ as 

 \begin{equation}\label{eq68}
{\mathrm{\Xi }}_0=b^2_2-3b_1b_3+12b_0b_4,\ \ \ {\mathrm{\Xi }}_1=2b^3_2-9b_1b_2b_3+27b_0b^2_3+27b^2_1b_4-72b_4b_2b_0.
\end{equation}

The numerical experiments on the quartic equation $b_4{\mathrm{\Omega }}^4+b_3{\mathrm{\Omega }}^3+b_2{\mathrm{\Omega }}^2+b_1\mathrm{\Omega }+b_0=0$ suggest that the possible conditions to get positive real roots are

 \begin{equation}\label{eq69}
\left\{{\mathrm{\Xi }}_0>0\wedge \left|\frac{{\mathrm{\Xi }}_1}{2\sqrt{{\mathrm{\Xi }}^3_0}}\right|<1\ \right\}\mathrm{\ }\mathrm{\wedge }\ \left\{b_2<0\ \mathrm{\vee }\ b_1<0\right\}.
\end{equation}

Combination of the above possible conditions \eqref{eq69} with the discriminants \eqref{eq68} and coefficients (\ref{eq67}) lead to the existence of a critical value ${\omega }_{nce}$ for the natural frequency such that when 

 \begin{equation}\label{eq70}
{\omega }_n>{\omega }_{nce}>{\omega }_{nc},
\end{equation}
there will be a positive real root $\mathrm{\Omega }={\check{\mathrm{\Omega }}}_e$ at which the amplitude $A_{\mathrm{\Omega }e}\left(\mathrm{\Omega }\right)$ is maximum. The values of this critical frequency for thermal resonance of the electron temperature are determined for copper, gold and lead in Table \ref{tabel4}, where the electron temperature may resonate in response to an external excitation frequency $\mathrm{\Omega }={\check{\mathrm{\Omega }}}_e$ provided that the natural frequency of the thermal system exceeds such a critical value \eqref{eq70}. 

On the other hand, the second equation of \eqref{eq65} and the frequency-dependent amplitude \eqref{eq63} lead to the sixth order (bicubic) equation for $\mathrm{\Omega }$:

 \begin{equation}\label{eq71}
d_3{\mathrm{\Omega }}^6+d_2{\mathrm{\Omega }}^4+d_1{\mathrm{\Omega }}^2+d_0=0,
\end{equation}
where the coefficients $d_3$, $d_2$, $d_1$ and $d_0$ are given by

 \begin{eqnarray}\label{eq72}
 	\nonumber
 	d_3=2{\tau }^2_T{\tau }^2_{q1},\ \ \ d_2=3{\tau }^2_T+{\tau }^2_{q1}\left[{\left(\frac{{\tau }_T}{{\tau }_{q1}}+\frac{{\tau }_T}{{\tau }_{q2}}\right)}^2-2{\tau }^2_T\left(\frac{1}{{\tau }_{q1}{\tau }_{q2}}+{\omega }^2_n\right)\right], \\
 	\nonumber
 	d_1=2\left[{\left(\frac{{\tau }_T}{{\tau }_{q1}}+\frac{{\tau }_T}{{\tau }_{q2}}\right)}^2-2{\tau }^2_T\left(\frac{1}{{\tau }_{q1}{\tau }_{q2}}+{\omega }^2_n\right)\right], \\
    d_0={\tau }^2_T{\left(\frac{1}{{\tau }_{q1}{\tau }_{q2}}+{\omega }^2_n\right)}^2-2\left(\frac{{\tau }_T}{{\tau }_{q1}}+\frac{{\tau }_T}{{\tau }_{q2}}\right){\omega }^2_n-{\tau }^2_{q1}{\omega }^4_n.
\end{eqnarray}

Again, the process of finding the positive real root of \eqref{eq71} is not a tractable process. Because all the coefficients of \eqref{eq71} are given near room temperature, Table \ref{tabel1}, except the natural frequency ${\omega }_n$, therefore determining the value of natural frequency at which equation \eqref{eq71} owns at least one positive real root is an indispensable step. In the same way we formulate the conditions \eqref{eq69} for the quartic equation, one can use a similar inductive numerical-based approach to deduce that the possible conditions for the bicubic equation \eqref{eq71} to have at least one positive real root are 

 \begin{equation}\label{eq73}
\left\{d^2_2-3d_1d_3>0\wedge \left|\frac{2d^3_2-9d_1d_2d_3+27d_0d^2_3}{2\sqrt{{\left(d^2_2-3d_1d_3\right)}^3}}\right|<1\right\}\ \mathrm{\wedge }\ \left\{d_2<0\ \mathrm{\vee }\ d_1<0\right\}.
\end{equation}

It is clear that the first two inequalities of \eqref{eq73} lead to the existence of real roots, while the third or the fourth inequality occurs when the root is positive real. Checking inequalities \eqref{eq73} with the coefficients (\ref{eq72}) for the sample of metals, we find that there is a critical natural frequency ${\omega }_{ncl}$ such that when 

\begin{equation}\label{eq74}
{\omega }_n>{\omega }_{ncl}>{\omega }_{nc},
\end{equation}
there exists a positive real root $\mathrm{\Omega }={\check{\mathrm{\Omega }}}_l$ for the external excitation frequency derived in \eqref{eq71} at which the frequency-dependent amplitude $\left|A_{\mathrm{\Omega }l}\left(\mathrm{\Omega }\right)\right|$ is maxima and the lattice temperature resonates. The values of ${\omega }_{ncl}$ for copper, gold and lead are approximated numerically in Table \ref{tabel4}. 

\begin{table}
	\centering
	\caption{Critical values of natural frequencies for the thermal resonance of electron temperature \textit{${\omega }_{nce}$} and lattice temperature \textit{${\omega }_{ncl}$} computed numerically near temperature  \textit{$T_0=300\ K$}.}
	\label{tabel4}
	\scalebox{0.9}{
\begin{tabular}{||cccc||} \hline \hline 
\textbf{Metal} &$\omega_{nc} \ (\mathrm{THz})$&$\omega_{nce} \ (\mathrm{THz})$&$\omega_{nc1} \ (\mathrm{THz})$ \\
\hline \hline 
\textbf{Cu} & 18.62 & $1.12846\times {\omega }_{nc}\ $ & $2.61948\times {\omega }_{nc}$ \\  \hline 
\textbf{Au} & 14.06 & $1.12619\times {\omega }_{nc}$ & $2.58165\times {\omega }_{nc}$ \\ \hline 
\textbf{Pb} & 113.78 & $1.12197\times {\omega }_{nc}$ & $2.52266\times {\omega }_{nc}$ \\ 
\hline \hline 
\end{tabular}
}
\end{table}

It is notable from Table \ref{tabel4} that the critical frequencies for thermal resonance of the electron temperature and the lattice temperature are approximately $1.13$ and $2.6$ of the critical frequency of underdamped temperature oscillations (${\omega }_{nce}\cong 1.13\ {\omega }_{nc}$ and  ${\omega }_{ncl}\cong 2.6\ {\omega }_{nc}$), namely, according to inequalities \eqref{eq70} and \eqref{eq74}, the thermal resonance is a high mode phenomenon which requires that the thermal system oscillates with underdamping mode. Furthermore, the critical frequency for electron temperature resonance is less than that for lattice temperature resonance, i.e.,

\begin{equation}\label{eq75}
{\omega }_{nc}<{\omega }_{nce}<{\omega }_{ncl}.
\end{equation}

\begin{figure}
	\centering
\includegraphics*[scale=0.35,angle=0]{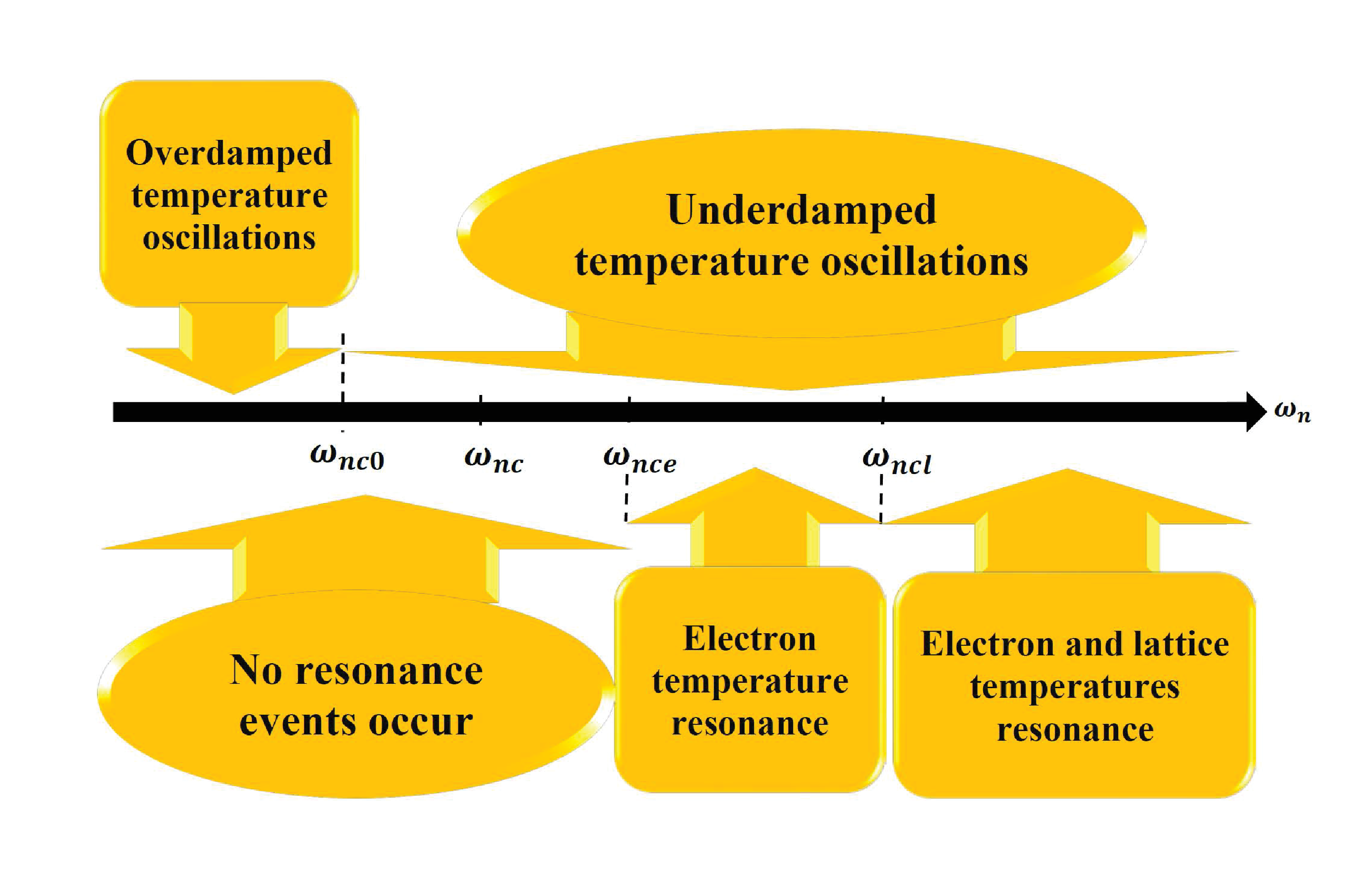}
\caption{ Schematic description of the critical frequency domains for electron and lattice thermal resonance. }
\label{fig4}
\end{figure}

This indicates to the presence of three critical frequency domains for the natural frequency containing three different events, see Fig. \ref{fig4}: (i) In ${\omega }_n<{\omega }_{nce}$, there are no resonance events; (ii) In ${\omega }_{nce}<{\omega }_n<{\omega }_{ncl}$, only the electron temperature may resonate; (iii) In ${\omega }_n>{\omega }_{ncl}$, both electron and lattice temperatures may resonate. This changes the classical concept of thermal resonance where there exist two frequency domains. Indeed, the external excitation frequency can lead thermal system governed by the one-step model if ${\mathrm{d}{\left|A_{\mathrm{\Omega }}\left(\mathrm{\Omega }\right)\right|}^2}/{\mathrm{d}\mathrm{\Omega }}=0$, where $A_{\mathrm{\Omega }}\left(\mathrm{\Omega }\right)$ is given by \eqref{eq64}. This condition yields the classical result that when ${\omega }_n>{0.6436}/{{\tau }_{q1}}$ \cite{RN35}, there will be a positive real external excitation frequency 

\begin{equation}\label{eq76}
 {\mathrm{\Omega }}_{max}=\sqrt{-\frac{1}{{\tau }^2_{q1}}+{\omega }_n\sqrt{{\omega }^2_n+\frac{2}{{\tau }^2_{q1}}}}.
\end{equation}

\begin{figure}
		\centering
\includegraphics*[scale=0.6,angle=0]{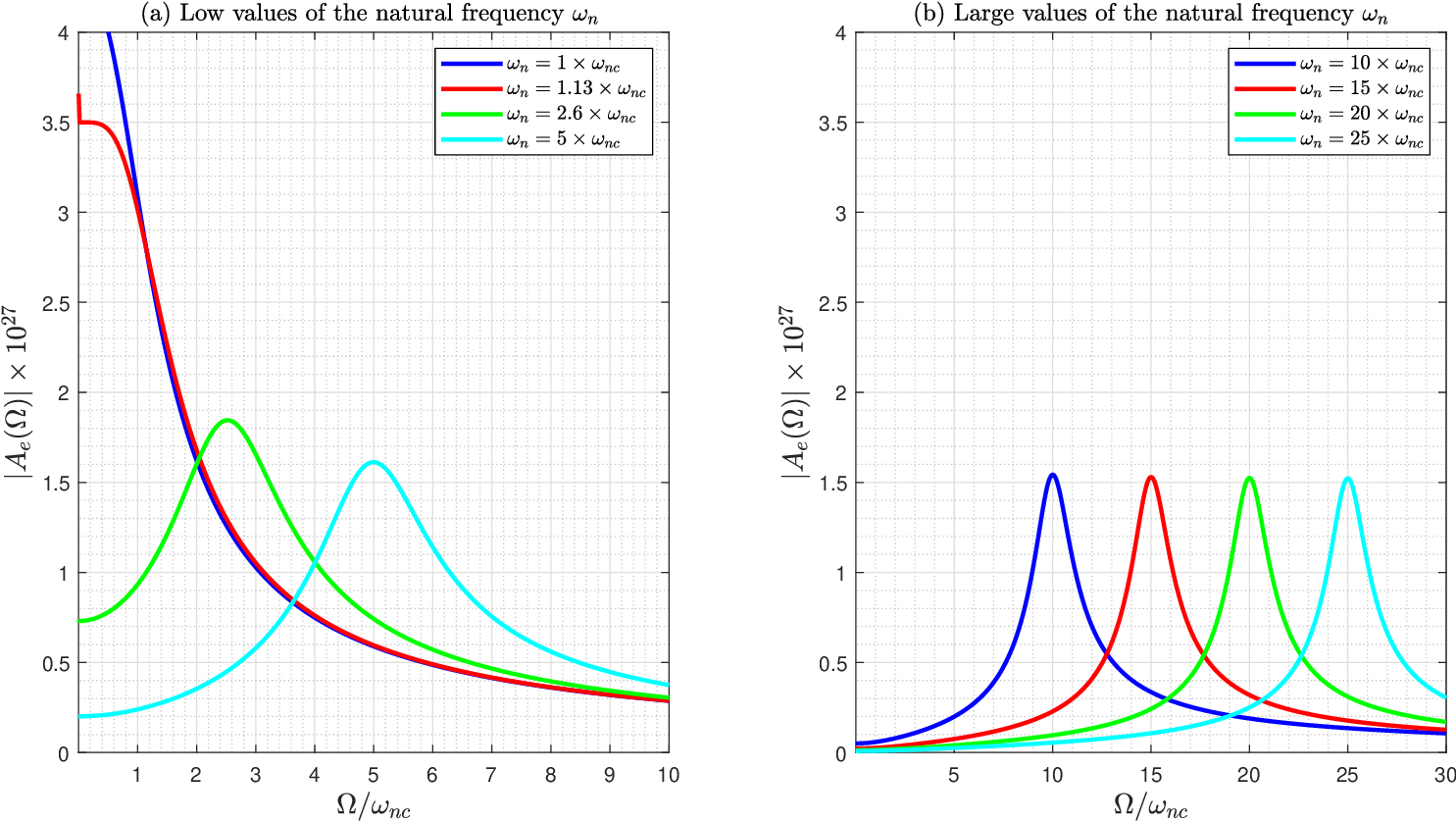} 
\caption{Frequency-Dependent amplitude of gold electron temperature \textit{$\left|A_{\mathit{\Omega}e}\left(\mathit{\Omega}\right)\right|$} at different values of possible natural frequency \textit{${\omega }_n$}. (a) \textit{${\omega }_n=1$}, \textit{$1.13$}, \textit{$2.6$}, \textit{$5$}; (b) \textit{${\omega }_n=10$}, \textit{$15$}, \textit{$20$}, \textit{$25$}.}
\label{fig5}
\end{figure}

\begin{figure}
			\centering
\includegraphics*[scale=0.6,angle=0]{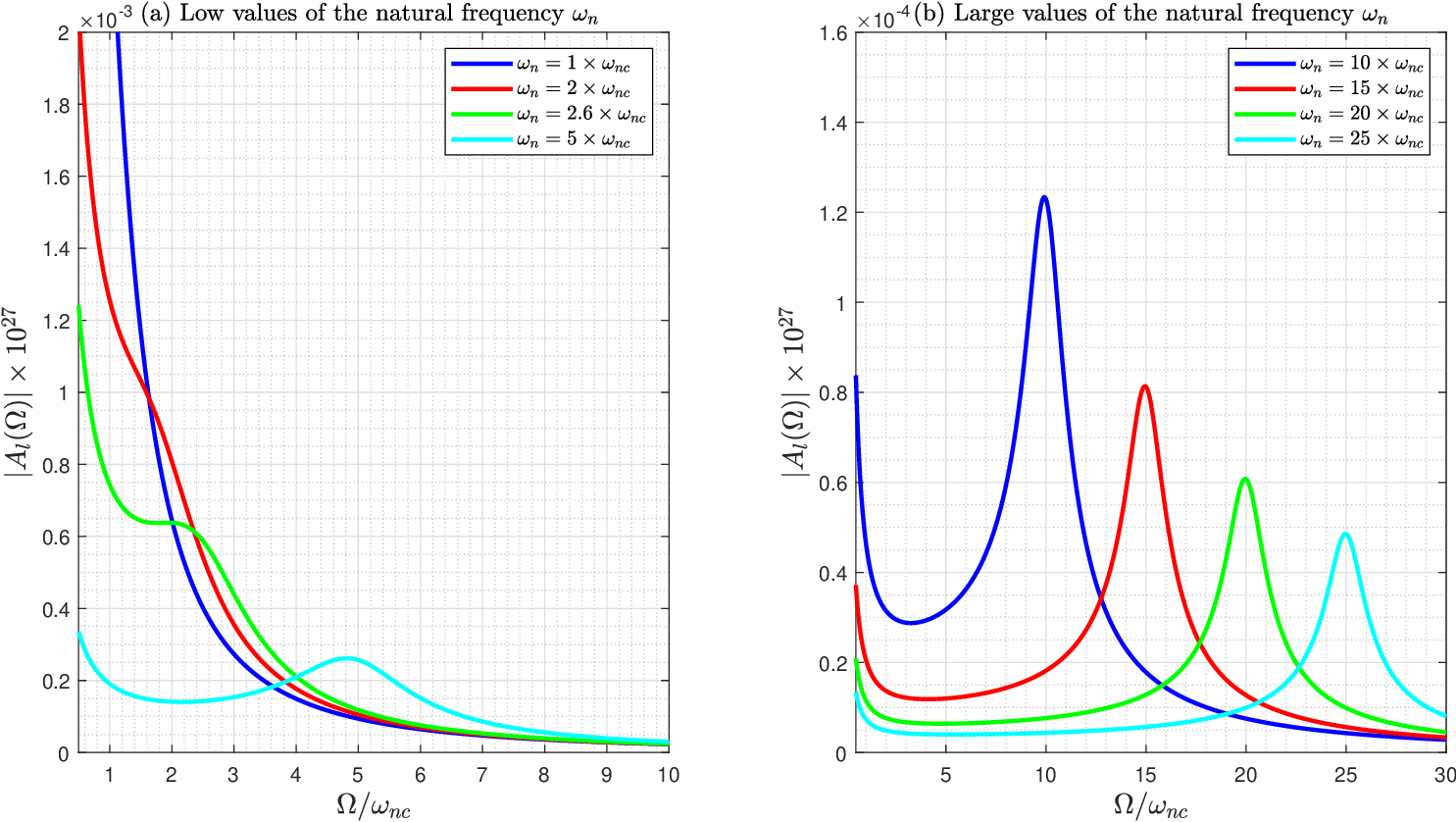}
\caption{Frequency-Dependent amplitude of gold lattice temperature \textit{$\left|A_{\mathit{\Omega}l}\left(\mathit{\Omega}\right)\right|$} at different values of possible natural frequency \textit{${\omega }_n$}: (a) \textit{${\omega }_n=1$}, \textit{$2$}, \textit{$2.6$}, \textit{$5$}; (b) \textit{${\omega }_n=10$}, \textit{$15$}, \textit{$20$}, \textit{$25$}.}
\label{fig6}  
\end{figure}

The absolute values of the frequency-dependent amplitudes, $\left|A_{\mathrm{\Omega }e}\left(\mathrm{\Omega }\right)\right|$ and $\left|A_{\mathrm{\Omega }l}\left(\mathrm{\Omega }\right)\right|$, given by

\begin{eqnarray}\label{eq77}
	\nonumber
	\left|A_{\mathrm{\Omega }e}\left(\mathrm{\Omega }\right)\right|=\sqrt{\frac{\left(1+{\tau }^{\mathrm{2}}_{q1}{\mathrm{\Omega }}^{\mathrm{2}}\right)\left(1+{\tau }^{\mathrm{2}}_T{\mathrm{\Omega }}^{\mathrm{2}}\right)}{{\left[{\omega }^2_n-{\tau }_T\left(\frac{1}{{\tau }_{q1}}+\frac{1}{{\tau }_{q2}}\right){\mathrm{\Omega }}^{\mathrm{2}}\right]}^2+{\tau }^{\mathrm{2}}_T{\mathrm{\Omega }}^{\mathrm{2}}{\left(\frac{1}{{\tau }_{q1}{\tau }_{q2}}+{\omega }^2_n-{\mathrm{\Omega }}^{\mathrm{2}}\right)}^2}}, \\
\left|A_{\mathrm{\Omega }l}\left(\mathrm{\Omega }\right)\right|=\sqrt{\frac{1+{\tau }^{\mathrm{2}}_{q1}{\mathrm{\Omega }}^{\mathrm{2}}}{{\left[{\omega }^2_n-{\tau }_T\left(\frac{1}{{\tau }_{q1}}+\frac{1}{{\tau }_{q2}}\right){\mathrm{\Omega }}^{\mathrm{2}}\right]}^2+{\tau }^{\mathrm{2}}_T{\mathrm{\Omega }}^{\mathrm{2}}{\left(\frac{1}{{\tau }_{q1}{\tau }_{q2}}+{\omega }^2_n-{\mathrm{\Omega }}^{\mathrm{2}}\right)}^2}},
\end{eqnarray}
are represented for gold versus the external excitation frequency $\mathrm{\Omega }$ in Fig. \ref{fig5} and Fig. \ref{fig6}, respectively, at different values of the natural frequency ${\omega }_n$. In Fig. \ref{fig5} (a), the values ${\omega }_n=1$ and $1.13$ are representatives to the domain ${\omega }_n<{\omega }_{nce}$ and the critical frequency ${\omega }_n\cong {\omega }_{nce}$. There is no a clear peak at these frequencies, however, the curve begins to be wrapped in preparation to record amplitude peaks later for the natural frequencies higher than ${\omega }_{nce}$. In the domain ${\omega }_n>{\omega }_{nce}$, there is a clear peak for the amplitude at $\mathrm{\Omega }={\mathrm{\Omega }}_e$, which refers to a thermal resonance of the electron temperature. The peak height becomes fixed ($\left|A_{\mathrm{\Omega }e}\left(\mathrm{\Omega }\right)\right|=1.525\times {10}^{-27}$) for large values of natural frequency at ${\mathrm{\Omega }}_e={\omega }_n$, see Fig. \ref{fig5} (b). This peculiarity was  mentioned by Tzou for the frequency-dependent amplitude \eqref{eq64} of the one-step model . Because ${\omega }_{nce}<{\omega }_{ncl}$, we see that whereas the electron temperature resonates at ${\omega }_{nce}<{\omega }_n\le {\omega }_{ncl}$, the lattice temperature may not be amplified in the vicinity of ${\omega }_{ncl}\cong 2.6{\omega }_{nc}$, refer to Fig. \ref{fig6} (a). When the natural frequency exceeds ${\omega }_{ncl}\cong 2.6{\omega }_{nc}$, the peaks can be observed in the frequency dependent amplitude at $\mathrm{\Omega }={\mathrm{\Omega }}_l$. For large values of the natural frequency, the amplitude peak is located at ${\mathrm{\Omega }}_l={\omega }_n$. Unlike the electron temperature amplitude, the peak heights of $\left|A_{\mathrm{\Omega }l}\left(\mathrm{\Omega }\right)\right|$ decrease with the increase of ${\omega }_n$, refer to Fig. \ref{fig6} (b). 

For natural frequencies less than the critical value ${\omega }_{ncl}$, the six roots of equation \eqref{eq71} are all complex, indicating that there are no external frequencies $\mathrm{\Omega }={\mathrm{\Omega }}_l$ can lead the lattice temperature to resonate. For the values ${\omega }_n>{\omega }_{ncl}$, on the other hand, equation \eqref{eq71} has two complex roots, two negative real roots, and two positive real roots. Since $\mathrm{\Omega }$ must be positive real, we exclude the complex and the negative real roots. We find that the two positive real roots of \eqref{eq71} have the mathematical expressions:
\[\ \ \ \] 

\begin{eqnarray}\label{eq78}
	\nonumber
	{\mathrm{\Omega }}_{l0}=\frac{2\sqrt{d^2_2-3d_1d_3}{\sin \left(\frac{1}{3}{{\sin}^{-1} \left(\frac{2d^3_2-9d_1d_2d_3+27d_0d^2_3}{2\sqrt{{\left(d^2_2-3d_1d_3\right)}^3}}\right)\ }\right)\ }}{3\left|d_3\right|}-\frac{d_2}{3d_3}, \\
{\mathrm{\Omega }}_l=\frac{2\sqrt{d^2_2-3d_1d_3}{\cos \left(\frac{1}{3}{{\cos}^{-1} \left(-\frac{2d^3_2-9d_1d_2d_3+27d_0d^2_3}{2\sqrt{{\left(d^2_2-3d_1d_3\right)}^3}}\right)\ }\right)\ }}{3\left|d_3\right|}-\frac{d_2}{3d_3},
\end{eqnarray}
where

\begin{equation}\label{eq79}
{\mathrm{\Omega }}_{l0}\le {\mathrm{\Omega }}_l.
\end{equation}

Because the positive real roots (\ref{eq78}) satisfy the relation ${\mathrm{d}\left|A_{\mathrm{\Omega }l}\left(\mathrm{\Omega }\right)\right|}/{\mathrm{d}\mathrm{\Omega }}=0$, therefore ${\mathrm{\Omega }}_{l0}$ and ${\mathrm{\Omega }}_l$ are critical values which lead to maximum value (thermal resonance) or minimum value to $\left|A_{\mathrm{\Omega }l}\left(\mathrm{\Omega }\right)\right|$. In fact Fig. \ref{fig6} shows that there are two critical point on the curve, one is a maximum (crest) locating at $\mathrm{\Omega }={\mathrm{\Omega }}_l$, and the other is a minimum (trough) locating at $\mathrm{\Omega }={\mathrm{\Omega }}_{l0}$. 

For electron temperature, one can numerically check that in the case ${\omega }_n<{\omega }_{nce}$, equation \eqref{eq66} has eight complex roots, whereas the case ${\omega }_n>{\omega }_{nce}$ gives four complex roots, two negative real roots and two positive real roots for equation \eqref{eq66}. Due to the complexity of the biquartic equation, we cannot guarantee the presence of these two real positive roots for the whole frequency domain ${\omega }_n>{\omega }_{nce}$, nor even a fixed specific form for these roots. For the natural frequency domain $\mathrm{1.12}<{{\omega }_n}/{{\omega }_{nc}}<40$, the relation between the excitation frequency (which leads to resonance) and the natural frequency can be approximated as \cite{RN45,RN46,RN47}

\begin{equation}\label{eq80}
{\mathrm{\Omega }}_e=\sqrt{-\frac{b_3}{4b_4}+S_1+\frac{1}{2}\sqrt{-4S^2_1-2p-\frac{q}{S_1}}},
\end{equation}
where 

\begin{eqnarray}\label{eq81} 
	\nonumber
	S_1=\frac{1}{2}\sqrt{-\frac{2}{3}p+\frac{1}{3b_4}\left(Q_1+\frac{{\mathrm{\Xi }}_0}{Q_1}\right)},\quad Q_1=\sqrt[3]{\frac{{\mathrm{\Xi }}_1+\sqrt{{\mathrm{\Xi }}^2_1-4{\mathrm{\Xi }}^3_0}}{2}},\ \ \ p=\frac{8b_2b_4-3b^2_3}{8b^2_4}, \\
	\nonumber
	q=\frac{b^3_3-4b_2b_3b_4+8b_1b^2_4}{8b^3_4},\ \ \ {\mathrm{\Xi }}_0=b^2_2-3b_1b_3+12b_0b_4, \\
{\mathrm{\Xi }}_1=2b^3_2-9b_1b_2b_3+27b_0b^2_3+27b^2_1b_4-72b_0b_2b_4.
\end{eqnarray}

\begin{figure}
			\centering
\includegraphics*[scale=0.5,angle=0]{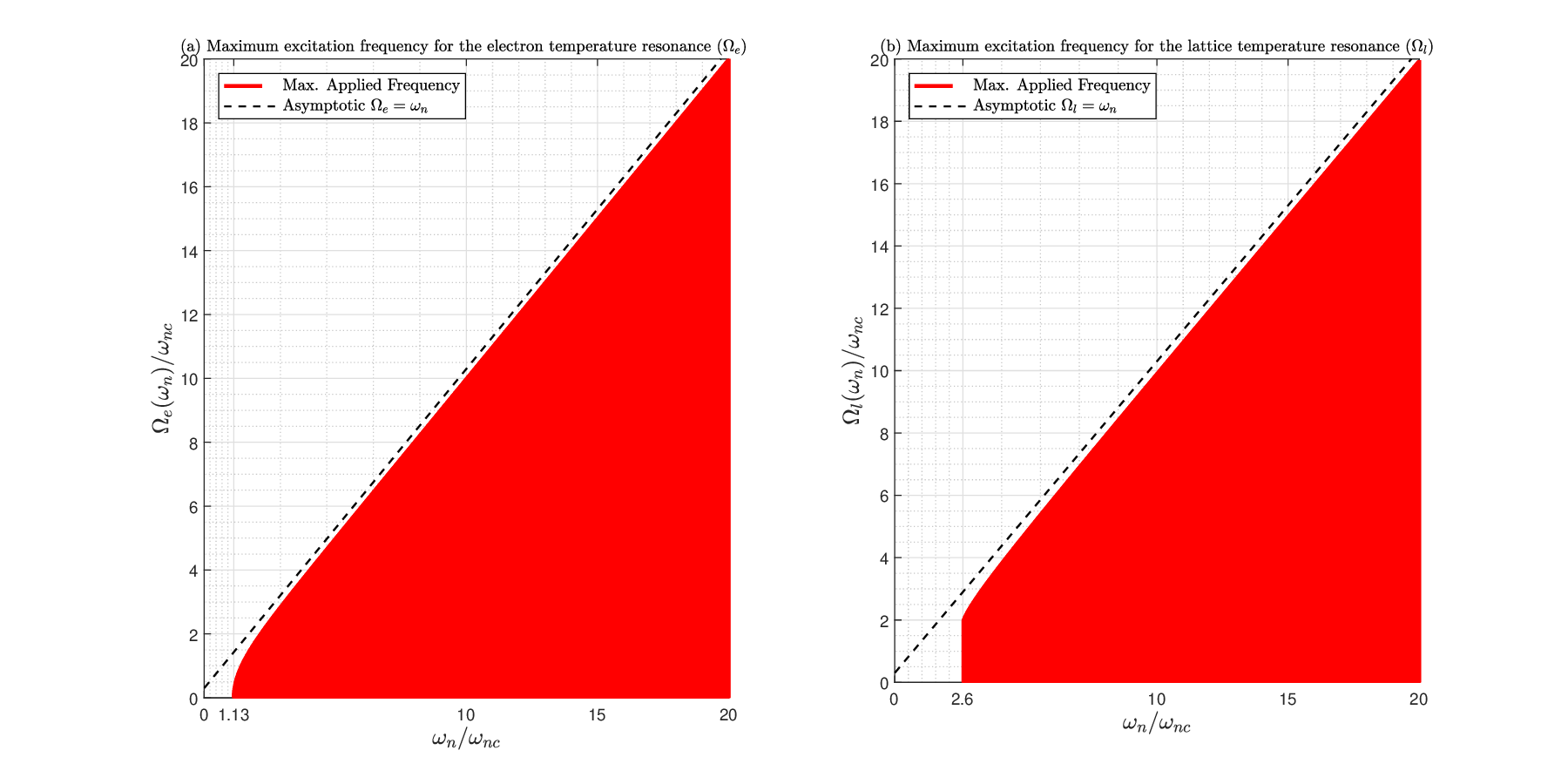}
\caption{Excitation frequency driving gold thermal system to resonate: (a) for electron temperature; (b) for lattice temperature.}
\label{fig7}  
\end{figure}

Otherwise, the form \eqref{eq80} may change its structure or sign such that it is not suitable to be an excitation frequency. In Fig. \ref{fig7}, we compare between the maximum applied excitation frequency for electron temperature ${\mathrm{\Omega }}_e$, given through \eqref{eq80}-(\ref{eq81}), and the maximum applied excitation frequency for the lattice temperature ${\mathrm{\Omega }}_l$, given by \eqref{eq78}. We can easily observe that both excitation frequencies exist provided that the natural frequency ${\omega }_n$ exceeds the value ${\omega }_{nce}=1.13\ {\omega }_{nc}$ for the electron temperature and the value ${\omega }_{ncl}=2.6\ {\omega }_{nc}$ for the lattice temperature. As the natural frequency increases sufficiently above the critical value ${\omega }_{ncl}$, both applied frequencies ${\mathrm{\Omega }}_e$ and ${\mathrm{\Omega }}_l$ (which drive the thermal system to resonance) become equal to ${\omega }_n$, then both temperatures may resonate together with different amplitude and decaying rates.

\section{ Conclusions}
\label{sec5}
In this work, we have studied the thermal oscillations of the electron and the lattice temperatures governed by the hyperbolic two-step model with one relaxation in the electron heat flux. The lattice heat conduction of the metal is disregarded throughout our analysis. It is found that when the natural frequency exceeds a critical value, the temperature oscillations become underdamped. The presence of a critical damped solution separating the underdamped and the overdamped oscillations is deduced by defining a second critical frequency. We have analytically derived this critical frequency for the underdamped thermal oscillations, \eqref{eq39}, such that for all natural frequency values greater than this critical frequency, ${\omega }_n>{\omega }_{nc}$, the underdamped thermal oscillations can be obviously observed. We have also determined an adjacent frequency domain ${\omega }_{nc0}<{\omega }_n<{\omega }_{nc}$, in which the characteristic roots are complex, but the underdamping is not effective. The lower critical frequency ${\omega }_{nc0}$ yields the critically damped solution, while the frequency domain ${\omega }_n<{\omega }_{nc0}$, results in the overdamped thermal oscillations. 

Thermal resonance phenomenon in the electron and lattice temperatures is examined. We have emphasized the classical result for the one-step model that the thermal resonance in the two-step model is a high-mode phenomenon in the sense that it occurs provided that the natural frequency of the thermal system exceeds a critical value. We have therefore defined two additional critical values ${\omega }_{nce}$ for the electron temperature and ${\omega }_{ncl}$ for the lattice temperature, such that for ${\omega }_n>{\omega }_{nce}$, the electron temperature can resonate when the external applied heat source is modulated at a frequency $\mathrm{\Omega }={\mathrm{\Omega }}_e={\omega }_n$, and for ${\omega }_n>{\omega }_{ncl}$, the lattice temperature can resonate when the external applied heat source is modulated at a frequency $\mathrm{\Omega }={\mathrm{\Omega }}_l={\omega }_n$. Because ${\omega }_{nce}<{\omega }_{ncl}$, we have three frequency domains with three different events. In the frequency domain ${\omega }_n<{\omega }_{nce}$, both electron and lattice temperature cannot be led to resonance under any external excitation frequency. In the frequency domain ${\omega }_{nce}<{\omega }_n<{\omega }_{ncl}$, the electron temperature may resonate if the external frequency is $\mathrm{\Omega }={\mathrm{\Omega }}_e$. In the third frequency domain ${\omega }_n>{\omega }_{ncl}$, both electron and lattice temperatures may resonate under the external excitation $\mathrm{\Omega }={\mathrm{\Omega }}_l$ which roughly equal to the natural frequency ${\omega }_n$ at high values.

\section*{Acknowledgment}

\noindent S.L. Sobolev was supported by the State Task of Russian Federation, State Registration AAAA-A19-119071190017-7

\bibliographystyle{RS}
\bibliography{resonance}

\end{document}